\documentclass[twocolumn,showpacs,preprintnumbers,amsmath,amssymb,superscriptaddress]{revtex4}

\usepackage{graphicx}
\usepackage{subfigure}

\usepackage{dcolumn}
\usepackage{bm}

\usepackage{color} 

\begin{document}


\title{A review of beam tomography research at Daresbury Laboratory}

\author{K. M. Hock} \email{kmhock@liverpool.ac.uk}
\address{Department of Physics, University of Liverpool, Liverpool L69 7ZE, United Kingdom}
\address{Cockcroft Institute, Daresbury Laboratory, Warrington WA4 4AD, United Kingdom}
\author{M. G. Ibison }
\address{Department of Physics, University of Liverpool, Liverpool L69 7ZE, United Kingdom}
\address{Cockcroft Institute, Daresbury Laboratory, Warrington WA4 4AD, United Kingdom}
\author{D. J. Holder}
\address{School of Physics and Astronomy, University of Manchester, Oxford Road, Manchester, M13 9PL, United Kingdom}
\address{Cockcroft Institute, Daresbury Laboratory, Warrington WA4 4AD, United Kingdom}
\author{B. D. Muratori}
\address{ASTeC, STFC, Daresbury Laboratory, Warrington WA4 4AD, United Kingdom}
\address{Cockcroft Institute, Daresbury Laboratory, Warrington WA4 4AD, United Kingdom}
\author{A. Wolski}
\address{Department of Physics, University of Liverpool, Liverpool L69 7ZE, United Kingdom}
\address{Cockcroft Institute, Daresbury Laboratory, Warrington WA4 4AD, United Kingdom}

\date{\today}


\begin{abstract}

This is a review on beam tomography research at Daresbury.  The research 
has focussed on development of normalised phase space
  techniques.  It starts with the idea of sampling tomographic projections at 
  equal phase advances and shows that this would give the optimal reconstruction 
  results.  This idea has influenced the design, construction and operation of 
  the tomography sections at the Photo Injector Test Facility at Zeuthen (PITZ)
  and at the Accelerator and Laser in Combined Experiments (ALICE) at Daresbury.
  The theoretical justification of this idea is later developed through 
  simulations and analysis of the measurements results at ALICE.  The 
  mathematical formalism is constructed around the normalised phase space and 
  the idea of equal phase advances become the basis of this.  This formalism 
  is applied to a variety of experimental and simulated situations and shown 
  to be useful in improving resolution, increasing reliability and providing 
  diagnostic information.   In this review, we also present the 
  simplifying concepts, formalisms and simulation tools that we have developed.

\end{abstract}

\maketitle


\tableofcontents 
\section{\label{sec:intro}Introduction}

Phase space tomography \cite{Stratakis, McKee} is a measurement technique that is used in 
accelerators to characterise the phase space of a particle beam.  It has 
been used in a number of accelerators, including PITZ \cite{Asova2}, 
UMER \cite{Stratakis2}, SNS, PSI \cite{Reggiani}, 
CERN \cite{CERN}, BNL \cite{Yakimenko}, FLASH \cite{Honkavaara} and
TRIUMF \cite{Rao2}.  
The beam distribution measured in
 coordinate space can be mapped mathematically to a phase space, and the rotation
 angle in the phase space can be varied by changing the strengths of optical
 elements along the beamline.  This mapping to rigid rotation makes it possible
 to reconstruct the phase space distribution using standard tomographic 
 techniques.  

In a simple implementation, the optical element could just be a drift space. 
 Suppose that we wish to determine the transverse, horizontal phase space at a
 particular location in a beamline.  Suppose that there is a scintillating 
 screen at a second location further along the beamline.  The horizontal phase 
 space at the screen is related to that at the first location.  Assuming linear 
 mapping, the relation can be represented by a matrix.  This matrix produces a 
 geometrical transformation on the phase space, usually a combination of 
 shearing and stretching.  The tomographic method involves projecting the 
 screen image on the horizontal axis. The geometric connection means that 
 this can be related to the projection of the phase space at the first 
 location in a rotated direction.  This angle can be varied by changing the
 length of the drift space.  By measuring the projections for a range of
 drift distances, the projections for a range of angles at the first location
 can be obtained.  The phase space distribution can be reconstructed from 
 these projections using techniques like Filtered Back Projection (FBP) or
 Maximum Entropy Technique (MENT).  In practice, the setup will involve a
 combination of drift spaces and other elements, such as quadrupoles or 
 solenoids.  Measurements on longitudinal phase space also require RF 
 cavities.  In this review we focus on transverse phase space.

Phase space tomography has been implemented in different ways in a number of
 accelerators.  At ALICE \cite{Ibison}, it follows closely the basic theory 
 described above.  It uses a quadrupole to change the rotation angle by changing the
 quadrupole strength.  The quadrupole strength is varied using a computer, and 
 screen images are captured automatically and then processed using FBP.  However,
 the range of angles accessible by a single quadrupole is limited to a smaller 
 range than the full 180$^\circ$.  At PITZ \cite{Asova2}, the rotation angle is varied
 using a combination of drift spaces and quadrupoles.  In practice it may not be
 easy to build a setup to move a screen mechanically along a beampipe in order 
 to vary the drift distance.  The PITZ tomography section consists of four 
 screens to measure the beam distribution in coordinate space at different 
 locations along the beampipe.  The quadrupole strengths are not normally varied
 during a measurement.  This means that only four projections are measured.  This
 small number of projection angles makes it better to use MENT for
 reconstruction.  At ALICE, PSI and SNS \cite{Reggiani}, the tomography diagnostic 
 sections have also been designed for MENT with between three and five screens. 
 At TRIUMF \cite{Rao2}, a wire scanner with a quadrupole 
 is used instead of screens.   There are wires in 
 three fixed angles.  These measure the projections in three directions in the
 transverse coordinate space.  MENT is used for reconstruction.   At UMER \cite{Stratakis2},
 the strengths of a few quadrupoles are adjusted to obtain the full 180$^\circ$ 
 range of angles.  The reconstruction is carried out using FBP.  The
 reconstruction algorithm is modified to include space charge effects.   

Phase space tomography research at ALICE over the past three years has focussed 
on two main areas:
 development of the normalised phase space method, and more recently development
 of 4D reconstruction.   Development of the normalised phase space method has 
 been primarily motivated by the idea of using equal phase advances in phase
 space tomography \cite{Asova}.  The phase advance here refers to betatron phase advance \cite{Lee}.  
There at four screens separated by FODO cells. This setup is designed 
 to give 45$^\circ$ phase advance in between adjacent screens in transverse phase
 space in both horizontal and vertical directions.  Together, the four screens 
 give four projections at equal phase advances over 180$^\circ$.  This design 
 has been adopted for the construction of the PITZ tomography section and used 
 in tomographic measurements since then \cite{Asova2}.  The same idea has been used in 
 the design and construction of the ALICE tomography section 
 \cite{Muratori, Muratori2}.  There are
 three screens with FODO cells in between adjacent screens.  Phase advance 
 between adjacent screens is adjusted to be 60$^\circ$.  This gives three
 projections with equal phase advances in between.  However, at ALICE we have 
 not had the chance to carry out such a measurement.  In the beam time 
 available, we have mainly used a quadrupole scan in which the
 quadrupole strength is varied rapidly using a computer and images captured at a
 single screen.  

At PITZ where only four projections are available, MENT is used for 
reconstructing the phase space.  The use of such a small amount of measured data 
to reconstruct the whole phase space means that the magnitude of error is
 uncertain.  Simulations on hypothetical distributions at PITZ have demonstrated
 that using equal phase advances give the smallest error in emittances of 
 reconstructed distributions \cite{Asova1}.  However, at the time of the construction of
 first PITZ and then ALICE, there has been no theoretical justification as to 
 why equal phase advances should produce optimal reconstructions.  This has been 
 a question because there is no obvious connection between phase advance and the 
 method of phase space tomography.  The only angle that exists in phase space
 tomography is the projection angle and this is not equal to phase advance.  The
 explanation comes later when we show that the phase advance is in fact equal to
 the projection angle in normalised phase space \cite{Hock}.  
 A Gaussian distribution in normalised phase space would appear roughly circular. 
 Having equal phase advances means sampling projections at equal angles in this 
 phase space.  This would be the natural sampling interval, particularly if 
 variation of distribution with angle is small.  Conversely, a distribution in 
 real phase space tends to be long and narrow because of long drift spaces in
 beamlines.  Such a distribution varies strongly with angle.  Using equal angle 
 intervals would either sample too little in the sharply varying directions or 
 require too many projections over the full 180$^\circ$ range.

This realisation has not only provided a theoretical justification for the idea
 and use of equal phase advances.  It also opens up new areas of applications. 
 So far, we have shown that normalised phase space can improve resolution for FBP
 \cite{Hock}, reduce distortion for MENT \cite{Hock1}, and detect linear errors in reconstructions 
 \cite{Ibison}.  
 In section \ref{sec:tomography}, we summarise the main steps
leading to the formulae for tomography in general.
 In section \ref{sec:phasespacetomography}, we provide a simple derivation of phase space tomography 
 formalism that we have developed.  
 In section \ref{sec:MENT}, we provide a simple derivaton of MENT.  
 In section \ref{sec:FBP}, we explain how we use  FBP in practice.
 In section \ref{sec:equal}, we review the idea
 of equal phase advances and how it has influenced the design 
 of tomography sections at PITZ and ALICE.
 In section \ref{sec:normalised},
 we explain our normalised phase space method and show how phase advance is 
 connected to projection angles.  
 In section \ref{sec:ALICE}, we summarise our measurement
 procedure at ALICE and the reconstruction results.  
 In section \ref{sec:space}, we explain an
 idea to observe space charge at the ALICE tomography section using normalised 
 phase space.  
 In section \ref{sec:MENT_recon}, we review the use of normalised phase space to
 improve the reliability of MENT reconstructions.  
 %
 %
 In section \ref{sec:conclude}, we conclude with a summary and 
 a discussion on what we plan to do next. 


\section{\label{sec:tomography}Tomography}


 \begin{figure}
\begin{center}
\includegraphics[width=.35\textwidth]{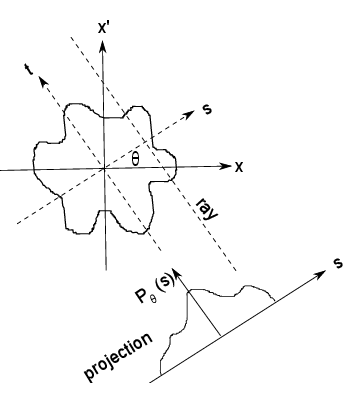}
\end{center}
\caption{\label{fig:tomography} 		 
}
\end{figure}

We review here the basic theory of tomography.  The goal is essentially
to derive a formula to calculate a 2D distribution function $f(x,x')$ from its 
projections.  The following steps are summarised from \cite{Kak}.  

We first define the projection.  Consider the axes $(s,t)$ rotated 
by angle $\theta$ in Fig. \ref{fig:tomography}.  
The coordinates are related to $(x,x')$ by
\begin{align}
  s &= x\cos\theta + x'\sin\theta \nonumber \\
  t &=-x\sin\theta + x'\cos\theta
\label{eq:ts}  
\end{align}
The projection of $f(x,x')$ along $s$ axis is given by
\begin{equation}
  P_\theta (s) = \int_{-\infty}^\infty f(x(s,t),x'(s,t)) dt
\end{equation}
This is an integral along a line of constant $s$.  This line is
called a ray.  It is perpendicular to the $s$ axis, which is the
direction of the projection.

The Fourier transform of the projection is
\begin{equation}
 S_\theta(w) = \int_{-\infty}^\infty P_\theta(s) e^{-i2\pi ws} ds
\label{eq:S} 
\end{equation}
Substituting the definition of projection
\begin{equation}
 S_\theta(w) = \int_{-\infty}^\infty 
               \left[ \int_{-\infty}^\infty f(x,x') dt \right] 
                e^{-i2\pi ws} ds
\end{equation}
and transforming to $(x,x')$ coordinates
\begin{equation}
  S_\theta(w) = \int_{-\infty}^\infty \int_{-\infty}^\infty
           f(x,x') e^{-i2\pi w(x\cos\theta + x'\sin\theta)} dx dx'
\end{equation}
In terms of the following coordinates
\begin{align}
  u &= w\cos\theta \nonumber \\
  v &= w\sin\theta
\label{eq:uv}  
\end{align}
we see that $S_\theta(w)$ is the same as the 2D Fourier transform
\begin{equation}
  F(u,v) = \int_{-\infty}^\infty \int_{-\infty}^\infty
           f(x,x') e^{-i2\pi (xu + x'v)} dx dx'
\label{eq:FFT}
\end{equation}
So
\begin{equation}
  S_\theta(w) = F(w,\theta) = F(w\cos\theta, w\sin\theta)
\label{eq:SF}
\end{equation}
where $(w,\theta)$ are the polar coordinates in the 2D spatial frequency
domain.

Inverting the transform gives 
\begin{equation}
 f(x,x') = \int_{-\infty}^\infty  \int_{-\infty}^\infty
          F(u,v) e^{i2\pi (ux+vx')} dudv
\label{eq:FFT_F}		  
\end{equation}
Transforming to polar coordinates:
\begin{equation}
  f(x,x') = \int_0^{2\pi} \int_0^\infty F(w,\theta)
        e^{i2\pi w(x\cos\theta + x'\sin\theta)} w dw d\theta
\end{equation}
Next split this into two parts:
\begin{align}
  &f(x,x') 
  = \int_0^{\pi} \int_0^\infty F(w,\theta)
        e^{i2\pi w(x\cos\theta + x'\sin\theta)} w dw d\theta \nonumber \\
  &+ \int_0^{\pi} \int_0^\infty F(w,\theta+\pi)
        e^{i2\pi w[x\cos(\theta+\pi) + x'\sin(\theta+\pi)]} w dw d\theta
\end{align}
Then use this property of Fourier transform:
\begin{equation}
  F(w,\theta+\pi) = F(-w,\theta)
\end{equation}
and rewrite the transform as:
\begin{equation}
 f(x,x') = \int_0^\pi \left[ \int_{-\infty}^\infty F(w,\theta) |w|
   e^{i2\pi ws} dw \right] d\theta
\end{equation}
where
$ F(w,\theta)$ is the Fourier transform of the projection
$S_\theta(w)$:
\begin{equation}
 f(x,x') = \int_0^\pi \left[ \int_{-\infty}^\infty S_\theta(w) |w|
   e^{i2\pi ws} dw \right] d\theta
\label{eq:FBP}
\end{equation}
This provides the relation between projections and function $f(x,x')$.

The Filtered Back Projection technique for computing $f(x,x')$ from
the projections is obtained by defining
\begin{equation}
  Q_\theta(s) = \int_{-\infty}^\infty S_\theta(w) |w| e^{i2\pi ws} dw
\label{eq:F}
\end{equation}
Multiplying a Fourier transform $S_\theta(w)$ by a function $|w|$
of frequency and then inverting the transform is often called 
filtering.  Since $S_\theta(w)$ is the Fourier transform of the
projection, $Q_\theta(s)$ is called the filtered projection.

From Eq. (\ref{eq:FBP})
\begin{equation}
 f(x, x') = \int_0^\pi Q_\theta(s) d\theta
\label{eq:BP}
\end{equation}
This is like spreading $Q_\theta(s)$ back over the $(x,x')$
space and then summing up for all angles.  For this reason, 
$Q_\theta(s)$ is called the back projection.

In principle, the two equations above can be discretised and used
to reconstruct $f(x, x')$ numerically from the projections 
$P_\theta(s)$.  This reconstruction technique is called
Filtered Back Projection.

\section{\label{sec:phasespacetomography}Phase Space Tomography}


A standard derivation of the equations used in beam tomography is given in
\cite{McKee}.  
Rather than just summarising the results, we reproduce here an alternative
derivation which gives insight into the geometric nature of the 
method. 

 \begin{figure}
\begin{center}
\includegraphics[width=.45\textwidth]{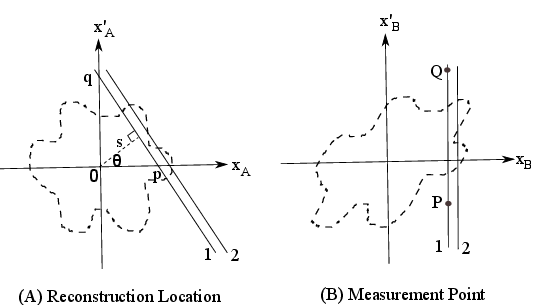}
\end{center}
\caption{\label{fig:projection} The $x$ intercept $a$ at A is mapped to
         point P at B, and $y$ intercept $b$ is mapped to point Q.
         Projection variable $s$ at A corresponds to projection 
         variable $x_B$ at B.		 
}
\end{figure}

We need to derive the relation between a projection at B in the
$x_B$ direction and the corresponding projection at A.  
Specifically, we want to find (i) a formula for the direction $\theta$ of
the projection at A; (ii) a formula to relate projection variables
$s$ and $x_B$; and (iii) a formula to relate a projection at B to
a projection at A.  
We assume that the mapping is given:
\begin{equation}
  \begin{pmatrix}
  x_B \\
  x_B'
 \end{pmatrix} =
  \begin{pmatrix}
  M_{11} & M_{12} \\
  M_{21} & M_{22}
 \end{pmatrix}
  \begin{pmatrix}
  x_A \\
  x_A'
 \end{pmatrix}
\label{eq:transfer}
\end{equation}
The effect of this mapping is a geometrical transformation.
For a drift space, it is a shear in the $x$ direction,
as illustrated in Fig. \ref{fig:projection}.  
For a thin quadrupole, it is a shear in the $x'$ direction.
For other elements, it could be some combination of shear, stretch and rotation.

Consider
a ray line 1 at B in Fig. \ref{fig:projection} and the corresponding
ray line 1 at A.  Line 1 at A is mapped to line 1 at B by the mapping
in Eq. (\ref{eq:transfer}).  The intercepts $p$ and $q$ at A are mapped
to points P and Q at B.  The coordinates of P are $(pM_{11}, pM_{21})$.
  The coordinates of Q are $(qM_{12}, qM_{22})$.  Since P and Q
  lie on the same vertical line, they have the same $x_B$ coordinate:
  \begin{equation}
    x_B = pM_{11} = qM_{12}.
	\label{eq:0pq}
  \end{equation}
From triangle 0pq in Fig. \ref{fig:projection}, angle $\theta$
is equal to angle 0qp.  So the tangent of the angle $\theta$ is $p/q$.
From Eq. (\ref{eq:0pq}), we obtain:
\begin{equation}
  \tan \theta = \frac{p}{q} = \frac{M_{12}}{M_{11}}.
  \label{eq:tangent}
\end{equation}
This gives the
formula for the direction $\theta$ of
the projection at A.

From Eq. (\ref{eq:0pq}), the ratio $x_B/s$ is equal to $pM_{11}/s$.
From triangle 0pq in Fig. \ref{fig:projection}, $p/s$ is equal to
$\sec\theta$.  Using the identity $1 + \tan^2\theta = \sec^2\theta$,  
Eq. (\ref{eq:tangent}) gives
\begin{align}
  \frac{x_B}{s} &= \frac{aM_{11}}{s} \nonumber \\
                &= M_{11}\sec\theta \nonumber \\
                &= M_{11}\sqrt{1 + \tan^2\theta} \nonumber \\
                &= M_{11}\sqrt{1 + \frac{M_{12}^2}{M_{11}^2}} \nonumber \\
                &= \sqrt{M_{11}^2 + M_{12}^2} 				
  \label{eq:scaling2}
\end{align}
This ratio is the scaling factor $a$
relating projection variables $s$ and $x_B$.

Compare the distance interval between lines 1 and 2 at B, and the 
corresponding interval at A.  The interval at A is clearly scaled down
 by the
above scaling factor $a$.  Since the number of particles within this
interval must be the same at A and at B, the projection $p_A$ at A
must be scaled up from the projection $p_B$ at B by $a$.  
This observation gives the formula to transform
a projection at B to a projection at A:  
\begin{equation}
  p_A(s) = a p_B(a s),
\label{eq:projection2}
\end{equation}
where $a$ is $x_B/s$.

This completes the derivation.
The 
 full set of equations needed to transform projections
from measurement point to reconstruction location
are:
\begin{equation}
  \tan \theta = \frac{M_{12}}{M_{11}}.
  \label{eq:tangent1}
\end{equation}
\begin{equation}
  a = \sqrt{M_{11}^2 + M_{12}^2}.				
  \label{eq:scaling2a}
\end{equation}
\begin{equation}
  s = \frac{x_B}{a}				
  \label{eq:scaling2b}
\end{equation}
\begin{equation}
  p_A = a p_B,
\label{eq:projection2a}
\end{equation}
After this transformation, each projection
at A corresponds to a simple rotation by angle $\theta$.

\section{\label{sec:MENT}Maximum Entropy Technique}


Our implementation of MENT follows closely the formalism described 
in \cite{Mottershead}.  We provide here a simpler derivation that 
allows us to replace most of the mathematical steps leading to
 the MENT equation with a pictorial explanation.

The initial steps in the MENT theory would be familiar to students of
physics who have studied the derivation of Boltzmann distribution.  
In a quantum system of particles, each particle can only
occupy discrete energy levels.  There are different arrangements
of particles that can give the same number at each level.
The Boltzmann distribution is the most likely distribution.
This is obtained by finding the distribution that has the greatest
number of arrangements.  Each arrangement must obey the constraints
that the total number of particles and the total energy are both
fixed.

\begin{figure}
\begin{center}
\includegraphics[width=.35\textwidth]{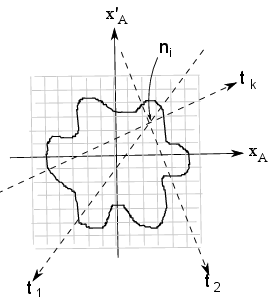}
\end{center}
\caption{\label{fig:ment1}The phase space is divided into a grid of tiny squares.
                          Each line labelled $t_k$ that goes through square $i$ is 
                          a ray that is parallel to the $t$ axis of projection angle $k$.						  
}
\end{figure}

In MENT, we divide a region of phase space into a grid of tiny 
squares as shown in Fig. \ref{fig:ment1}.  
Each square corresponds to an energy level.  A phase space
distribution tells us the number of particles in each square.
MENT aims to find the most likely distribution.  This is obtained
by finding the one with the largest number of possible arrangements.
The constraints are that the resulting distribution must 
give projections that agree with the measured ones at each angle.

We first review the mathematical steps leading to the Boltzmann
distribution \cite{Glazer}, and then show how this can be generalised directly to 
MENT.  Consider a system of $N$ distinguishable particles.  
Each particle can occupy the energy levels $\epsilon_i$, and there
are $n_i$ particles at each level.  The constraints are that the 
number of particles
\begin{equation}
  N = n_1 + n_2 + ...
\end{equation}
and the total energy
\begin{equation}
  U = n_1\epsilon_1 + n_2\epsilon_2 + ...
\end{equation}
are fixed.  A distribution is given by the set of numbers 
$(n_1, n_2, ...)$.  The number of possible arrangements for this 
distribution is:
\begin{equation}
  W = \frac{N!}{n_1!n_2!...}
\end{equation}
We want to find the distribution for which $W$ is maximum.  This
would be easier if we maximise $\ln W$ instead because Stirling's
approximation makes the factorials simpler:
\begin{equation}
  \ln W \approx (N\ln N - N) - (n_1\ln n_1 - n_1) - (n_2\ln n_2 - n_2) - ...
\end{equation}
In physics, entropy is given by $k_B \ln W$ where $k_B$ is Boltzmann's constant.
Hence the name Maximum Entropy Technique.
We then apply the method of Lagrange multipliers \cite{Glazer}.  First we
make the Lagrange function
\begin{equation}
  L = \ln W + \lambda_0 \sum n_i + \lambda_1 \sum n_i\epsilon_i
\end{equation}
where the second and third terms on the right come from the two
constraints above, and $\lambda_0$ and $\lambda_1$ are called 
Lagrange multipliers.  If we now maximise $L$ with respect to
$(n_1, n_2, ...)$, we would get the most likely distribution under
the given constraints.  First we differentiate and set the derivative
to zero:
\begin{equation}
  \frac{\partial L}{\partial n_i} = 
  -\ln n_i + \lambda_0 + \lambda_1 \epsilon_i = 0
\end{equation}
Then we solve for $n_i$:
\begin{equation}
  n_i = e^{\lambda_0} e^{\lambda_1 \epsilon_i}
\end{equation}
When $\lambda_1$ is replaced with $-1/k_BT$ using thermodynamic 
 reasoning, we get the familiar Boltzmann distribution.
 
 To apply this to MENT, we replace energy levels with the tiny squares
 in phase space.  The formula for the number of arrangements $W$ is the 
 same. The constraint on the number of particles $N$ is also the same.
 The constraint on total energy is now replaced by the constraints
 that the projections for the distribution must agree with the
 the measured ones:  
\begin{equation}
  p_k(s_k) = \sum_{t_k} n_i
\end{equation}
The left side of this equation is the projection value for the angle
$k$ and coordinate $s_k$.  The subscript $t_k$ of the summation on the
right side means that only those tiny squares on the ray 
at angle $k$ and coordinate $s_k$ are included in the sum.
The ray indicated by  $t_k$  is illustrated in Fig. \ref{fig:ment1}.  
The length of a ray in one square is different from its length in another
square.  The size of each square is assumed to be very small so that
the sum over $t_k$ approaches an integral.

The Lagrange function is then given by
\begin{equation}
  L = \ln W + \lambda_0 \sum n_i + \sum_{k} \sum_{s_k} \lambda_k(s_k) \sum_{t_k} n_i
\end{equation}
$s_k$ is assumed to be discrete with very small steps so that the sum over $s_k$
approaches an integral.  In Mottershead, the integral signs would be used for 
both the sumes over $s_k$ and $t_k$.  We retain the summation signs to simplify
the next step.

To maximise $L$, we differentiate with respect to $n_i$:
\begin{equation}
  \frac{\partial L}{\partial n_i} = 0 = 
  -\ln n_i + \lambda_0 + \sum_{k} \lambda_k(s_k)
\end{equation}
To understand the sum on the right, observe that all
$n$ variables should vanish except $n_i$ because the differentiation
is with respect to $n_i$.  Recall that $n_i$ is the population of
particles at a particular tiny square.  The multipliers $\lambda_k(s_k)$
that remain must correspond to those rays that pass through 
the centre of this particular square, as illustrated in Fig.
\ref{fig:ment1}. The number of rays is just the
number of projections.  $s_k$ would be the coordinate of the square's
centre for each projection.  With this understanding, we now rearrange
to get  
\begin{equation}
  n_i = e^{\lambda_0}\prod_{k=1}^K e^{\lambda_k(s_k)}
\end{equation}
where $K$ is the number of measured projections.
This can be rewritten as 
\begin{equation}
  f(x,x') = \prod_{k} h_k(s_k)
\end{equation}
where we have defined
\begin{equation}
  h_k(s_k) = e^{\lambda_k(s_k) + \lambda_0/K}
\end{equation}
We have equated the number density function $f(x,x')$ to 
the population $n_i$.  This is correct up to a constant
factor.

The key result is that the number density  
$f(x_A, x_A')$ of particles in 
phase space at the reconstruction
location A
can be expressed as a product of certain functions.  
Each of these functions
has only one variable, and this variable is the distance along each
projection direction $s$ (see Fig. \ref{fig:tomography}).  This relation
can be written as
\begin{equation}
  f(x_A, x_A') = \prod_{k=1}^K h_k(s_k(x_A, x_A')),
  \label{eq:product}
\end{equation}

 Recall the constraint that $f(x_A, x_A')$ must give the correct
 projection that has been measured for each projection.  The $k^{th}$ projection
 is related to $f(x_A, x_A')$ by
\begin{equation}
  p_k(s_k) = \int f(x_A, x_A') dt_k
  \label{eq:constraint}
\end{equation}
where $t_k$ is the axis perpendicular to the $s_k$ axis, and the integral is over 
the range of $t_k$ where $f(x_A, x_A')$ is nonzero.  
The coordinates $(x_A, x_A')$ are determined
for each value of $s_k$ given on the left of the equation and 
each value of $t_k$ defined during the 
 integration.  This means that if $f(x_A, x_A')$ is known, then when
we integrate it along the $t_k$ direction for a given $s_k$ value, the answer must be
equal to the value of the projection at $s_k$.

Equations (\ref{eq:product}) and (\ref{eq:constraint}) fully define
 the mathematical problem and 
the distribution $f(x_A, x_A')$ can in principle be solved.

Using Eqs. (\ref{eq:product}) and (\ref{eq:constraint}), we can now solve for 
the distribution $f(x_A, x_A')$.  By substituting Eq. (\ref{eq:product})
into Eq. (\ref{eq:constraint}), we get
\begin{equation}
  p_k(s_k) = h_k(s_k) \int dt_k \prod_{k' \not= k}^K h_{k'}(s_{k'}(x_A, x_A')) 
  \label{eq:product2}
\end{equation}
where $h_k(s_k)$ is factored out.  This is possible because $s_k$ and $t_k$ are the 
coordinates of the $k^{th}$ projection $p_k(s_k)$, so $s_k$ does not change when
$t_k$ is varied in the integral.
Equation (\ref{eq:product2}) makes it possible to solve for
the unknown $h_k(s_k)$ using a technique known as Gauss-Seidel iteration:
\begin{enumerate}  
\item  Rearrange Eq. (\ref{eq:product2}) for iteration:
\begin{equation}
  h_k^{i+1}(s_k) = \frac{p_k(s_k)}{ \int dt_k \prod_{{k'} \neq k} h_{k'}^i[s_{k'}(x_A, x_A')] }
  \label{eq:product2a}
\end{equation}
where $h_k^i(s_k)$ is the result for $h_k(s_k)$ after $i$ iterations.
\item   Use initial values of $h_k^0(s_k) = 1$.
\item  Use Eqs. (\ref{eq:product}) and (\ref{eq:constraint}) to calculate the 
        projections  for from the $i^{th}$ iteration:
\begin{equation}
 p_k^i(s_k) = \int dt_k  \prod_{{k'}=1}^K h_{k'}^i[s_{k'}(x_A, x_A')].
  \label{eq:product2b}
\end{equation}
\item  Calculate the differences between $p_k^i(s_k)$ and the measured $p_k(s_k)$
       for all $s_k$.
\item  Repeat the iteration until this difference is small enough for all $s_k$.
       (For the calculations in this paper, we stop when the difference 
        at each pixel $(x_A, x_A')$ is less than a tolerance level of
        1\% of the peak value of $f(x_A, x_A')$).
\end{enumerate}  

The computed projections may not always converge to the 
measured projections.  We have found that if the projections
are too noisy or if they remain non-zero up to the limits of the
domain of $s_n$, the method fails.  For the projections used in
this paper, convergence is usually achieved after 3 or 4 iterations.

\section{\label{sec:FBP}FBP in Practice}


In this section, we explain the main steps involved in processing 
measured projections.  We also describe simulation tools we have 
used to validate reconstruction codes.

\subsection{\label{sec:centroid}Centre of reconstruction}

\begin{figure}
\begin{center}
\includegraphics[width=.35\textwidth]{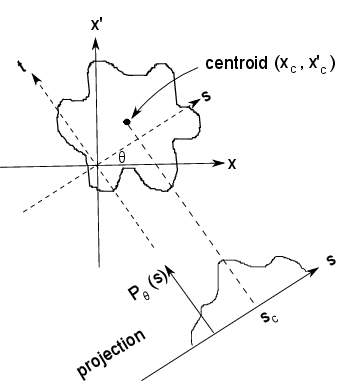}
\end{center}
\caption{\label{fig:centre}	A distribution in which the
                         centroid is not at the origin.					  
}
\end{figure}

In measured projections, there is a key information that is missing
for the reconstruction.  It is the origin.  In the derivaton of
the Filtered Back Projection technique, notice that each projection
has an origin.  Consider what happens if the origin of one projection
is in error.  Then during reconstruction, one of the back projections would
be shifted.  When this is added to other back projections,  the
result would clearly be erroneous.  Unfortunately, in measured
projections, we do not know where the origin of each projection is.
This is not an issue that is normally discussed in papers on phase 
space tomography.  Here, we describe our solution.  We shall prove that
if we take the centroid of each projection to be its origin, the resulting
reconstruction would be identical to the actual distribution.

Our solution is to use the centroid of each projection as the origin.

To show that this gives the correct reconstruction, consider a 
hypothetical distribution $f(x,x')$ in phase space.
Its centroid position is given by:
\begin{align}
  x_c &= \int \int x f(x,x') dx dx' \\
  x_c' &= \int \int x' f(x,x') dx dx'
\end{align}
Suppose that the centroid is not at the origin.
So the centroid of the projection $P_\theta(s)$ is also not at its origin
$s = 0$. Suppose that it is at $s_c$.
From Fig. \ref{fig:centre}, this is given by
\begin{align}
  s_c = x_c\cos\theta + x_c'\sin\theta 
\label{eq:t_c}  
\end{align}
 Suppose that we use this as the origin for reconstruction.
 The centroid can be determined directly from a measured projection
 using the centroid formula
\begin{align}
  s_c = \int s P_\theta(s) ds
\end{align}
This would return the same value using any arbitrary point as $s = 0$.
Taking $s_c$ as the origin of a projection, the new projection is
\begin{align}
  P_\theta'(s) = P_\theta(s - s_c)
\end{align}
If we now put this through the FBP equations, we first obtain the
Fourier transform of $P_\theta'(s)$.  From the property of
Fourier transform or from Eq. (\ref{eq:S}):
\begin{equation}
 S_\theta'(w) = e^{-i2\pi ws_c} S_\theta(w) 
\end{equation}
From Eq. (\ref{eq:SF}), we can write this as
\begin{equation}
 S_\theta'(w) = F'(w,\theta) = e^{-i2\pi ws_c} F(w,\theta) 
\end{equation}
Substituting into Eq. (\ref{eq:FFT_F}) gives
\begin{align}
 f'(x,x') &= \int_{-\infty}^\infty  \int_{-\infty}^\infty
          F'(u,v) e^{i2\pi (ux+vx')} dudv \\
         &= \int_{-\infty}^\infty  \int_{-\infty}^\infty
          e^{-i2\pi ws_c} F(u,v) e^{i2\pi (ux+vx')} dudv
\end{align}
Substituting Eqs. (\ref{eq:t_c}) and (\ref{eq:uv}) gives
\begin{equation}
 f'(x,x') = \int_{-\infty}^\infty  \int_{-\infty}^\infty
          e^{-i2\pi (ux_c+vx_c')} F(u,v) e^{i2\pi (ux+vx')} dudv
\end{equation}
Finally, comparing with Eq. (\ref{eq:FFT_F}) gives
\begin{equation}
 f'(x,x') = f(x-x_c, x'-x_c')
\end{equation}
This completes the proof that the distribution reconstructed using
the projection centroids as origins is identical to the actual
distribution $f(x,x')$, up to a rigid translation.

\subsection{\label{sec:intervals}Nonuniform angle intervals}

In the usual implementation of FBP, Eq. (\ref{eq:BP}) is discretised
as 
\begin{equation}
 f(x, x') = \frac{\pi}{K} \sum_{k=1}^K Q_{\theta_k}(x\cos\theta_k + x'\sin\theta_k)
\label{eq:BP_1}
\end{equation}
where Eq. (\ref{eq:ts}) is used to express $s$ in terms of $x$ and $x'$.
The angle interval given by $\pi/K$ is assumed to be uniform.
This is usually valid, for example in X-ray Computer Aided Tomography  
scan in which rotation angles can be precisely controlled.

In phase space tomography, it is convenient to allow the angle intervals
to be different.  The main reason is that the angle is varied by changing
the strengths of optical elements.  This variation need not be linear.
In the case of a quadrupole with a drift space for example, the 
variation can be a combination of steep and gentle. 
An analytic formula for projection angle in terms of quadrupole current is
available.  In principal, it should be possible to develop a
numerical code to compute precise values of currents required for 
uniform angles.
In practice, we obtain the currents from tabulated values of
currents and angles using linear interpolation. 
To check the accuracy of the currents obtained in this way, 
the analytic formula is used to compute the corresponding angles.
The results show that this procedure is prone to numerical errors.  
We may find that the actual intervals are not exactly uniform.  
We can then
correct for this by simply using the actual angle intervals in
the back projection equation.  So Eq. (\ref{eq:BP_1})
should be written as 
\begin{equation}
 f(x, x') = \sum_{k=1}^K Q_{\theta_k}(x\cos\theta_k + x'\sin\theta_k) \Delta\theta_k
\label{eq:BP_2}
\end{equation}
where $\Delta\theta_k$ is the actual angle interval.

Two other (hopefully) less common situations where this would be
useful are when there is a systematic error in the quadrupole current
or an error in beam energy at the stage of determining the required
currents.  Suppose that these errors are discovered after the
 measurements.  Because the projection angle does not vary linearly
 with current or energy, the resulting reconstruction could be
 completely wrong.  One option would be to redo the experiment.
But there is another option.  We can recompute the angles using the
corrected currents and energy.  The new angle intervals can then
be used in Eq. (\ref{eq:BP_2}) and the correct distribution computed.

\subsection{\label{sec:centroid}Hypothetical Gaussian distribution}

Whether it is to verify a reconstruction code written by others or to check
a code written by ourselves, it is useful to verify 
the reconstruction procedure using a hypothetical
distribution with known projections.  Reconstructing using these
projections must obviously return the original distribution.  If it does not,
then we know there is an error in the code.

In tomography in general, it is common to use a distribution made
up of ellipses of different shapes, sizes and brightness.  
An example is the Shepp Logan phantom which consists of ellipses
arranged to look like organs in a cross-section of a human
body.  There are simple formulae to compute the projections 
of the combination of ellipses \cite{Kak}.  In phase space tomography,
these ellipses with sharp edges do not look realistic.  Instead,
it is better to use Gaussian distributions.  Fortunately, we can
also derive analytic formulae for the projections of a Gaussian
distribution.  We would like to be able to describe the distribution
using Twiss parameters and compute the projection for a given
transfer matrix.  We list here the formulae that we have derived 
and used for the simulations shown in later sections.

Suppose that we need a Gaussian distribution at reconstruction location
A with emittance
$\epsilon$, beta function $\beta$ and alpha function $\alpha$.
The Gaussian distribution is given by
\begin{equation}
  f(x_A,x_A') = \exp\left(-\frac{x_N^2 + x_N'^2}{a_0^2}\right)
\end{equation}
where
\begin{equation}
 a_0 = \sqrt{\epsilon}
\end{equation}
and 
\begin{equation}
\left( \begin{array}{c} x_N \\ 
                       x'_N \end{array} \right) 
= \left( \begin{array}{cc} \frac{1}{\sqrt{\beta}}      & 0 \\ 
                  \frac{\alpha}{\sqrt{\beta}} & \sqrt{\beta} \end{array} \right) \cdot
  \left( \begin{array}{c} x_A \\ 
                          x_A' \end{array} \right).
\label{eq:norm}
\end{equation}
This is just the transformation to a normalised phase space.

Suppose that it is mapped to the screen
at location B
 where the horizontal projection is measured.  Suppose that the
 mapping is given by matrix $R_0$.  
 Then the distribution at B is
 \begin{equation}
  f(x_B,x_B') = \exp\left(-\frac{x_N^2 + x_N'^2}{a_0^2}\right)
\end{equation}
where
\begin{equation}
\left( \begin{array}{c} x_B \\ 
                       x'_B \end{array} \right) 
= R_0
  \left( \begin{array}{c} x_A \\ 
                          x_A' \end{array} \right).
\label{eq:AB}
\end{equation}
 
 The projection along $x_B$ axis is given by
\begin{equation}
 p(x_B) =  \int_{-\infty}^\infty   f(x_B,x_B') dx_B'
\end{equation}
Doing the integration gives 
\begin{equation}
 p(x_B) = a_0 \sqrt{A\pi} \exp\left(-\frac{Ax_B^2}{a_0^2}\right)
\end{equation}
where 
\begin{align}
  A &= (a^2 + c^2) - Be^2  \\
  e &= (ab + cd)/B         \\
  B &= b^2 + d^2          
\end{align}
\begin{equation}
\left( \begin{array}{cc} a      & b \\ 
                         c      & d \end{array} \right)  
 = N_1 R_0^{-1}
\end{equation}
 and $N_1$ is the matrix in Eq. (\ref{eq:norm}).
 

\section{\label{sec:equal}Equal Phase Advances}


As far as we can trace in the literature, the idea of using equal phase 
advances in phase space tomography may have originated from 
a simulation study on emittance measurement for the Tesla Test Facility
\cite{Castro}.  This empirical study shows that when 4 screens at 45$^\circ$
phase advances in a FODO lattice are used, the emittance computed using
 images from the 4 screens has the smallest error.  
 
The Tesla Test Facility design described in \cite{Castro} consists
of two diagnostic sections at two different locations.  
Both are intended for measuring emittance, not phase space tomography.  
Each section consists of 4 screens.  Between each pair of adjacent screens
is a FODO cell.  The three FODO cells form a short periodic structure.
The intention is to measure the beamwidths at the 4 screens.
Together with the transfer matrices of the FODO cells, the emittance
can then be calculated \cite{Castro}.  

In this study, the strengths of the FODO cells are optimised to reduce the 
errors of measurement.  The following procedure is adopted:  
\begin{enumerate}
\item  The FODO structure is assumed to be infinitely periodic.  
       So the beta functions are determined by the periodicity condtion.
\item  The phase advances are then computed.
       These would be equal between adjacent screens since the structure is periodic.
\item  In the actual beamline, the FODO structure is not infinitely periodic.
       So the actual magnets along the beamline must
       be adjusted to match the beam to the FODO struture before a measurement.
\end{enumerate}

A simulation is then carried out in \cite{Castro} to determine the performance.
This simulation determines how the error in measured emittance would vary 
with error in beam size measurement at each screen.  A random error is added to
the beam sizes and the emittance calculated.  This is repeated for 1000 times
and the RMS emittance error is determined.  This is then repeated for a number
of phase advances.  The result shows that the RMS emittance error is smallest when
phase advance between adjacent screens is 45$^\circ$.

 This result has provided the
 justification for the design of the PITZ tomography section \cite{Asova1}.
 This has a similar design as the diagnostic section in the Tesla Test Facility,
 with 4 screens and a FODO cell in between each pair of adjacent screens.
 This FODO structure is also designed to give 45$^\circ$ in between screens.  
 This design was developed by a collaboration between PITZ and Daresbury.  
 It has subsequently influenced the design of the ALICE tomography section.  
 
 The ALICE section
 has 3 screens and a FODO cell in between adjacent screens.  This is perhaps
the first active use of the idea of equal phase advances for phase space 
tomography.  Whereas the
PITZ choice of 45$^\circ$ phase advance is empirically justified by emittance
studies, no such study has been carried for ALICE.  The choice of 60$^\circ$ 
phase advance for the ALICE tomography section comes from dividing 180$^\circ$
by 3.  The ``180$^\circ$'' figure comes from the full angular range for tomographic
projections.  The result in \cite{Castro} that 45$^\circ$ phase advance
is optimal is associated conceptually with tomographic projection angles -
45$^\circ$ is 180$^\circ$ divided by 4, the number of screens.  
So at ALICE, 180$^\circ$ divided by 3 because there are 3 screens.
There is thus a conceptual leap from empirical emittance study
to phase space tomography.  

\begin{figure}
  \begin{center}
    \subfigure[]{\label{fig:ALICE_beta_a}
      \includegraphics[width=.4\textwidth]{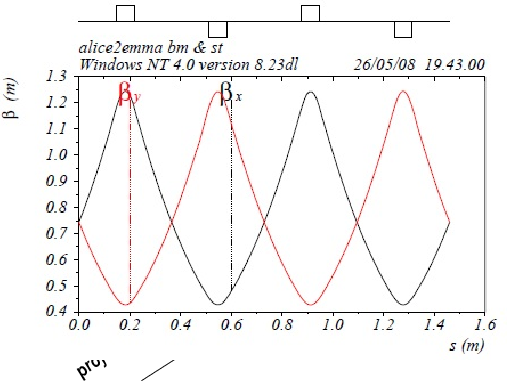}}
    \subfigure[]{\label{fig:ALICE_beta_b}
      \includegraphics[width=.4\textwidth]{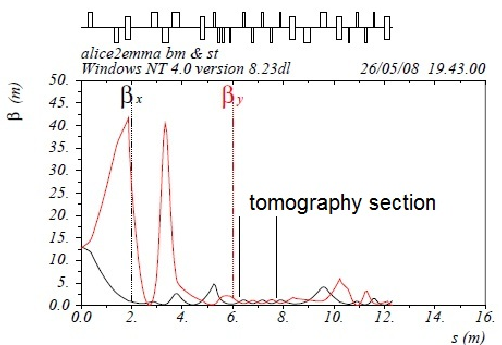}} 
  \end{center}
  \caption{(a) Beta functions at ALICE tomography section,
               assuming that the FODO lattice is periodic.  
           (b) Beta functions along beamline is matched into
               entrance of tomography section. }
  \label{fig:ALICE_beta}
\end{figure}

Figure \ref{fig:ALICE_beta_a} shows a schematic diagram of the
two FODO cells at the ALICE tomography section and the beta
functions computed using MAD8 under the assumption that the 
FODO cells are periodic \cite{Muratori}.
Figure \ref{fig:ALICE_beta_b} shows the lattice of ALICE magnets
before the tomography section and the beta functions of this lattice
that are matched using MADX to the periodic beta functions at the entrance to
the tomography section.

At the time of the construction of ALICE, there has 
been no further elaboration on this, whether it is emittance study or theoretical
analysis.  This comes later when we show in \cite{Hock} 
that phase advance is equal to projection
angle interval in normalised phase space.


\section{\label{sec:normalised}Normalised Phase Space}


In this section, we shall review the steps in \cite{Hock} to show that
phase advance is equal to projection angle interval in normalised phase
space.

Recall that
phase advance corresponds to rotation angles in the normalised phase space.
We assume that there is no coupling between vertical and horizontal motion
between reconstruction location and measurement point (e.g. screen).  This would
be true if we only use quadrupoles and drift spaces.
Then the horizontal, transverse normalised phase space at the reconstruction
location is defined by the following transformation:
\begin{equation}
\left( \begin{array}{c} x_N \\ 
                       x_N' \end{array} \right) 
= \left( \begin{array}{cc} \frac{1}{\sqrt{\beta}}      & 0 \\ 
                  \frac{\alpha}{\sqrt{\beta}} & \sqrt{\beta} \end{array} \right) 
  \left( \begin{array}{c} x_A \\ 
                          x_A' \end{array} \right).
\label{eq:actual-norm}
\end{equation}
$x_N$ and $x_N'$ are the corresponding co-ordinates in the normalised phase space,
and $\alpha$ and $\beta$ are Twiss parameters.  The Twiss parameters are determined 
by the second moments of the beam distribution:
\begin{align}
  \langle x^2 \rangle &= \beta\epsilon  \\
  \langle xx' \rangle &= -\alpha\epsilon \\
  \langle x'^2 \rangle &= \gamma\epsilon \\
  \epsilon &= \sqrt{\langle x^2 \rangle\langle x'^2 \rangle - \langle xx' \rangle^2}
\end{align}
A similar transformation to Eq. (\ref{eq:actual-norm}) applies to the vertical 
displacement $y$.
Reconstruction in normalised phase space can be done with a simple
extension of the method given in section \ref{sec:phasespacetomography}.
A matrix transforms the initial distribution at the reconstruction 
location to the distribution at the screen.  Based on this matrix,
the procedure in section \ref{sec:phasespacetomography} reconstructs the initial
 distribution.

The initial distribution may be considered the result of the transformation 
of the distribution in normalised phase space to real phase space.
The transformation is given by the inverse of Eq. (\ref{eq:actual-norm}).
In order to reconstruct in normalised phase space, we only need to replace
the matrix in Eq. (\ref{eq:transfer}), by a matrix that transforms the 
distribution all the way from the normalised phase space to the distribution 
at the screen.  This matrix is simply a product of the matrix in 
Eq. (\ref{eq:transfer}), and the matrix that transforms from normalised to 
real phase space.  The latter matrix may be obtained by inverting
Eq. (\ref{eq:actual-norm}) as follows:
\begin{equation}
  \left( \begin{array}{c} x_A \\ 
                          x_A' \end{array} \right)
= \left( \begin{array}{cc}  \sqrt{\beta_A}                & 0 \\ 
                  -\frac{\alpha_A}{\sqrt{\beta_A}} & \frac{1}{\sqrt{\beta_A}} \end{array} \right)
\left( \begin{array}{c} x_N \\ 
                       x_N' \end{array}  \right) 
\label{eq:norm-actual}
\end{equation}
where the subscript A means that the Twiss parameters refer to position A.
The matrix on the right hand side is the required matrix.
Inserting this into the right hand side of Eq. (\ref{eq:transfer})
gives the new transfer matrix $\tilde{M}$ needed for the reconstruction in 
normalised phase space:
\begin{equation}
  \tilde{M} = \left( \begin{array}{cc}
             M_{11} & M_{12} \\
             M_{21} & M_{22} 
             \end{array} \right)  
\left( \begin{array}{cc}  \sqrt{\beta_A}                & 0 \\ 
                  -\frac{\alpha_A}{\sqrt{\beta_A}} & \frac{1}{\sqrt{\beta_A}} \end{array} \right)
\label{eq:tomo-norm}
\end{equation}

We now demonstrate that the projection angle $\theta$ in the normalised
phase space is equal to the phase advance $\mu$.  This can be done using
the relation between the transfer matrix and the Twiss parameters at
positions A and B:
%
\begin{align}
&\left( \begin{array}{cc}
             M_{11} & M_{12} \\
             M_{21} & M_{22} 
             \end{array} \right) =  \nonumber \\
&\left( \begin{array}{cc}  \sqrt{\frac{\beta_B}{\beta_A}}(\cos\mu+\alpha_A\sin\mu)                                                         & \sqrt{\beta_B\beta_A}\sin\mu \\ 
                          \frac{\alpha_A-\alpha_B}{\sqrt{\beta_B\beta_A}}\cos\mu-\frac{1+\alpha_B\alpha_A}{\sqrt{\beta_B\beta_A}}\sin\mu  & \sqrt{\frac{\beta_A}{\beta_B}}(\cos\mu-\alpha_B\sin\mu) \end{array} \right) 
\label{eq:matrix-lattice-0}
\end{align}
%
where the subscript B means that the Twiss parameters refer to position B.
This can also be written as
%
\begin{align}
 \left( \begin{array}{cc}
             M_{11} & M_{12} \\
             M_{21} & M_{22} 
             \end{array} \right) =  
\left( \begin{array}{cc}  \sqrt{\beta_B}                    & 0 \\ 
                -\frac{\alpha_B}{\sqrt{\beta_B}} & \frac{1}{\sqrt{\beta_B}} \end{array} \right) 
& \left( \begin{array}{cc}  \cos\mu  &  \sin\mu \\ 
                -\sin\mu  &  \cos\mu \end{array} \right)  \nonumber \\
& \times \left( \begin{array}{cc}  \frac{1}{\sqrt{\beta_A}}        & 0 \\ 
                 \frac{\alpha_A}{\sqrt{\beta_A}} & \sqrt{\beta_A} \end{array} \right)
\label{eq:matrix-lattice}
\end{align}
%
We can understand the right hand side in a simple way:  the distribution at
A (reconstruction location) is transformed to normalised phase space, propagated to
B (screen) by a rigid rotation through angle $\mu$, and transformed back to
real phase space.
Substituting this into Eq.\,(\ref{eq:tomo-norm}), we find:
\begin{equation}
\tilde{M} = 
\left( \begin{array}{cc}  \sqrt{\beta_B}                    & 0 \\ 
                -\frac{\alpha_B}{\sqrt{\beta_B}} & \frac{1}{\sqrt{\beta_B}} \end{array} \right) 
\left( \begin{array}{cc}  \cos\mu  &  \sin\mu \\ 
                -\sin\mu  &  \cos\mu \end{array} \right)
\label{eq:tomo-norm-1}
\end{equation}
We can now apply Eq. (\ref{eq:tangent}) to this matrix
to find $\theta$.
Note that the original transfer matrix $R$ in Eq. (\ref{eq:transfer})
has been changed to $\tilde{M}$ defined in Eq.\,(\ref{eq:tomo-norm}).
So $M_{11}$ and $M_{12}$ in Eq. (\ref{eq:tangent})
must also be replaced by the elements in the first row of 
Eq. (\ref{eq:tomo-norm}).  These are equal to those in the
first row of Eq. (\ref{eq:tomo-norm-1}), which are
$\sqrt{\beta_B} \cos\mu$ and $\sqrt{\beta_B} \sin\mu$.
Substituting these into Eq.\,(\ref{eq:tangent}) for
$M_{11}$ and $M_{12}$ respectively, 
we find:
\begin{equation}
   \tan \theta = \tan \mu .
\label{eq:norm-angle}
\end{equation}
So  $\mu$ is indeed the projection angle.  

At this stage, we emphasise that the significant result is that if the
tomographic reconstruction is performed \emph{without} a normalising
transformation, then the projection angles need to be calculated 
from the transfer matrices: they are not simply the phase
advances.  This is significant for tomography at PITZ an ALICE,
which are designed with uniform betatron phase advance between
successive screens \cite{Holder,Loehl,Asova}, i.e. uniform distribution of projection angles in
\emph{normalised} phase space.  
The distribution of angles in real
phase space will not necessarily be uniform. This would have a direct impact
on the reconstruction.  

\begin{figure}
  \begin{center}
    \subfigure[]{\label{fig:actual-norm-a}
      \includegraphics[width=.22\textwidth]{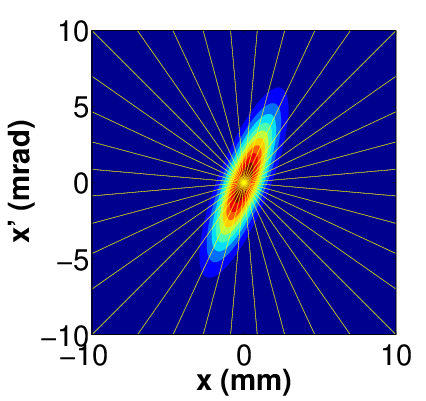}}
    \subfigure[]{\label{fig:actual-norm-b}
      \includegraphics[width=.22\textwidth]{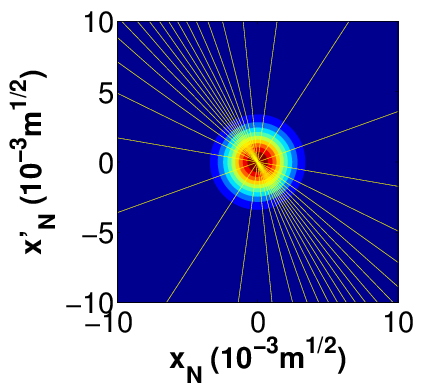}} 
  \end{center}
  \caption{(a) Real phase space with rays at uniform angular intervals.  
           (b) Normalised phase space.}
  \label{fig:actual-norm}
\end{figure}

\begin{figure}
  \begin{center}
    \subfigure[]{\label{fig:norm-actual-a}
      \includegraphics[width=.22\textwidth]{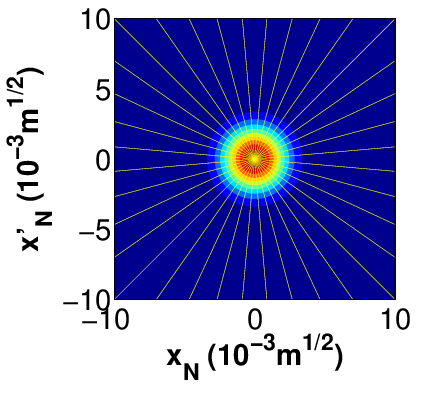}}
    \subfigure[]{\label{fig:norm-actual-b}
      \includegraphics[width=.22\textwidth]{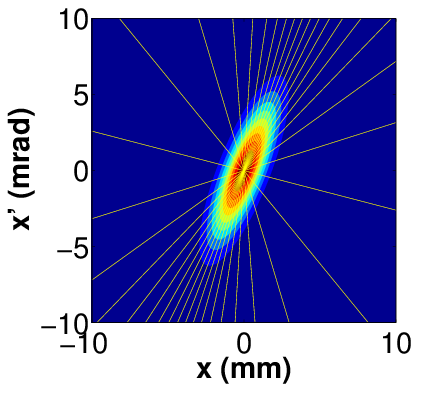}} 
  \end{center}
  \caption{(a) Normalised phase space with rays at uniform angular intervals.  
           (b) Real phase space.}
  \label{fig:norm-actual}
\end{figure}

To illustrate this point, consider 
the corresponding rays in real and normalised phase spaces shown
in Fig. \ref{fig:actual-norm}.  (The projection direction is perpendicular to the ray.)
Fig. \ref{fig:actual-norm-a} shows a Gaussian distribution in  real phase space,
with rays that are at uniform angular intervals.  In  normalised phase space, some
of the intervals become smaller, whereas others become larger, as shown in 
Fig. \ref{fig:actual-norm-b}. 
Fig. \ref{fig:norm-actual} illustrates the effect of the opposite
transformation -- starting with uniform intervals of angles in  normalised
phase space, shown in Fig. \ref{fig:norm-actual-a}.
This results in a nonuniform distribution of rays in  real phase space,
shown in Fig. \ref{fig:norm-actual-b}.  
These observations have direct impact on the reconstruction.
The actual effect depends on whether FBP or MENT is used.

\begin{figure}
  \begin{center}
    \subfigure[]{\label{fig:hotspots-actual-a}
      \includegraphics[width=.22\textwidth]{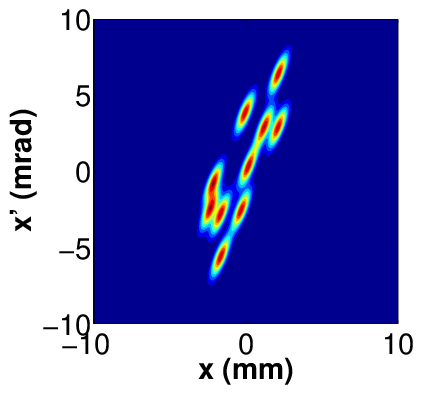}}
    \subfigure[]{\label{fig:hotspots-actual-b}
      \includegraphics[width=.22\textwidth]{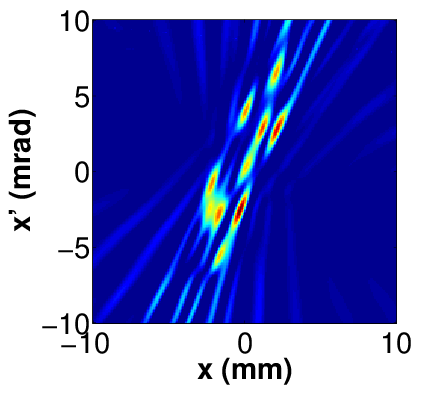}} 
  \end{center}
  \caption{(a) Distribution in real phase space.    
           (b) Reconstruction in real phase space. }
  \label{fig:hotspots-actual}
\end{figure}

Beam distributions are often more complex than simple Gaussians.
We consider a more complex hypothetical case where the distribution 
is made up of a group of closely spaced Gaussian spots,
as shown in Fig. \ref{fig:hotspots-actual-a}.
This provides 
a test of the ability of a reconstruction method to resolve the spots.
Note that each spot has a circular distribution in normalised phase space.
Assume a hypothetical system of eighteen screens separated only by drift spaces.
The projection angle corresponding to each screen can be chosen by adjusting the
length of the drift space using Eq. (\ref{eq:tangent}).
Start with the case of equal angular intervals in 
real phase space.  The projections from the screens are used to
reconstruct Fig. \ref{fig:hotspots-actual-a}. The result is shown in 
Fig.\,\ref{fig:hotspots-actual-b}.  The spots are all reproduced and at
 the correct positions.  However the resolution is less
clear.

\begin{figure}
  \begin{center}
    \subfigure[]{\label{fig:hotspots-norm-a}
      \includegraphics[width=.22\textwidth]{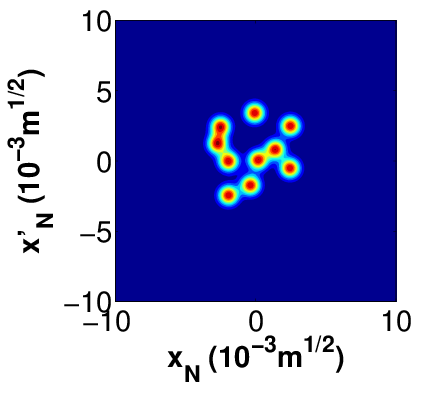}}
    \subfigure[]{\label{fig:hotspots-norm-b}
      \includegraphics[width=.22\textwidth]{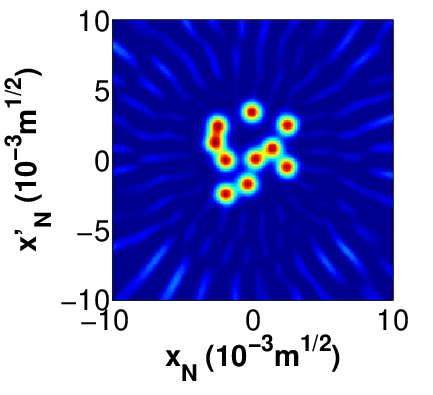}} 
  \end{center}
  \caption{(a) Distribution in normalised phase space.    
           (b) Reconstruction in normalised phase space. }
  \label{fig:hotspots-norm}
\end{figure}

\begin{figure}
\begin{center}
\includegraphics[width=.22\textwidth]{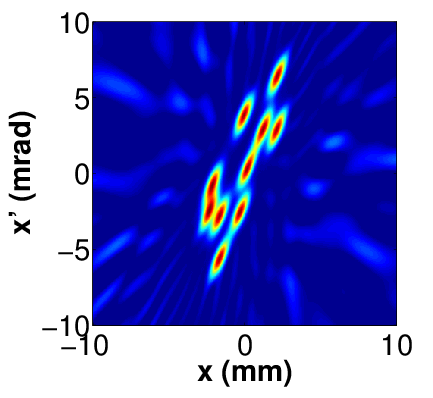}
\end{center}
\caption{\label{fig:stretch} 
		   Distribution obtained by transforming
         the co-ordinates in the reconstruction in normalised phase space
		 to the real phase space.
		 }
\end{figure}

We then look at the case of equal angular intervals in 
normalised phase space.
When Fig. \ref{fig:hotspots-actual-a} is transformed to normalised 
phase space, the distribution is as shown in 
Fig. \ref{fig:hotspots-norm-a}.
Note that the screens would now be at different positions from the previous case.
When we use the projections from these screens to
reconstruct the distribution in normalised phase space, we get
Fig. \ref{fig:hotspots-norm-b}.  This time, the spots are clearly reproduced.
The obvious step to transform the co-ordinates to real phase space 
gives Fig. \ref{fig:stretch}.  This is much clearer than
Fig. \ref{fig:hotspots-actual-b}.  Apart from the 
faint artefacts, the spots look almost the same as 
the original Fig. \ref{fig:hotspots-actual-a}.

One way to transform from Fig. \ref{fig:hotspots-norm-b} to Fig. 
\ref{fig:stretch} is to make a square grid of pixel positions 
for Fig. (\ref{fig:stretch}), compute the corresponding
positions in Fig.  \ref{fig:hotspots-norm-b} using Eq. (\ref{eq:actual-norm}),
then interpolate using the reconstructed Fig. \ref{fig:hotspots-norm-b}.
But there is a more direct way in which we can avoid the interpolation error.
Recall that a reconstruction is computed using Eq. (\ref{eq:BP}).
Instead of using this to compute Fig. \ref{fig:hotspots-norm-b} first, 
we can use this to compute the distribution at coordinates in normalised
phase space that correspond to the square grid in Fig. \ref{fig:stretch}.
In this way, Fig. \ref{fig:stretch} can be obtained directly
from the projections.

We should mention that to use this method for quadrupole scan, 
the Twiss parameters at the
reconstruction location must be measured first.  This can be done using a standard
method, e.g. as described in \cite{Ross} or \cite{Castro}.
From experience, we find that the method is quite robust.  For the
method to provide some benefit in reconstruction and the applications
described in the following sections, an estimate of the Twiss
parameters is often sufficient.


\section{\label{sec:ALICE}ALICE Tomography Section}


In this section we describe the experimental setup at the tomography diagnostic
 section in the ALICE-to-EMMA injection line that we use for our measurements.

\begin{figure}
\begin{center}
\includegraphics[width=.3\textwidth]{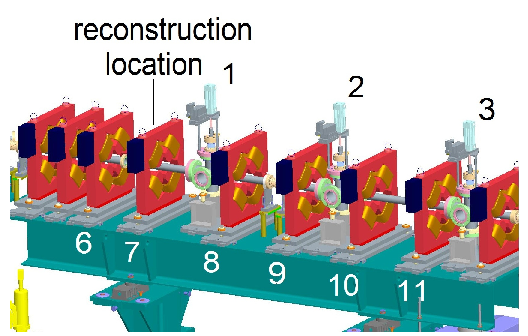}
\end{center}
\caption{\label{fig:tomo-section} ALICE tomography section:
           The tomography diagnostic section of the ALICE-to-EMMA injection line.
}
\end{figure}

The full-energy electron beam in ALICE is typically varied between 10 MeV (for 
injection into EMMA) to 27 MeV (for FEL operation). In our experiments, we have
 only used 12 MeV. The tomography section consists of three YAG screens, with two
 quadrupoles in between each adjacent pair of screens, as shown in Fig. 
 \ref{fig:tomo-section}. The
 three screens are labelled 1 to 3. The electron beam travels in the direction
 from screen 1 to screen 3. The distance from screen 1 to screen 3 is 1.5 m. The
 quadrupoles of interest are labelled 7 to 11. The length - between the entrance
 and exit planes - of each of these quadrupoles is 50 mm. The quadrupole scans for
 our experiments are carried out using quadrupoles 7 and 10. We shall refer to
 these as QUAD-07 and QUAD-10 respectively. The other quadrupoles are all fixed 
 at a current of 1.05~A during the scan.

\begin{figure}
\begin{center}
\includegraphics[width=.3\textwidth]{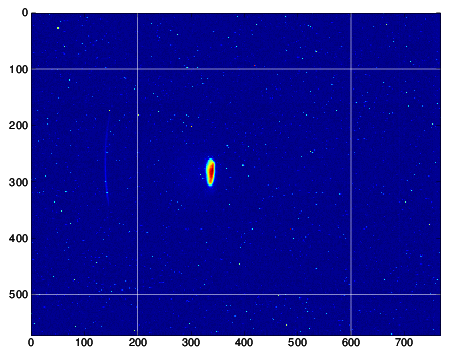}
\end{center}
\caption{\label{fig:beamimage}The image of the beam on screen 1. The ratio of 
                               distance on the screen to pixel size in the image is
							   0.0818 mm/pixel.
}
\end{figure}

Many factors influence the measurements. Figure \ref{fig:beamimage} shows an image, 
taken by a camera (Pacific Board Cameras PC-375 Mono, 752x582 pixels, 8 bit) 
focused on screen 1 in Fig. \ref{fig:tomo-section}, of a single bunch of charge of
 20 pC. The size and shape of this image can be adjusted by changing the strength 
 of QUAD-07, as well as all the other quadrupoles upstream of it. This is the 
 feature that is used in a quadrupole scan. The size and shape of the image is 
 also affected by day-to-day variation in the setup of ALICE, as well as 
 shorter-term instabilities. This can lead to variation of the image from bunch to
 bunch. A quadrupole scan or a tomographic reconstruction makes use of a set of 
 images, each taken at a slightly different time. The resulting emittance, Twiss 
 parameters or reconstructed phase space derived from these images must therefore 
 include some averaging of the bunch-to-bunch variations. Other variables include 
 the response of the YAG screen and the response of the camera. We assume that the
 intensity recorded by the camera image is directly proportional to the number of 
 electrons falling on each pixel.

For tomographic measurements, the transfer functions of the quadrupoles must be
 known accurately. This requires knowledge of the magnetic field gradient in each
 quadrupole. We rely on field gradient versus current measurements provided by the
 manufacturer. Note that there is hysteresis in the quadrupole magnets; thus the 
 field gradient can be slightly different, depending on the previous level of 
 excitation. The hysteresis curve provided by the manufacturer shows that at one 
 ampere current, the maximum error in the field is 7\%. This error remains constant 
 up to about 5~A, and is thus a potential source of measurement error.

The bunch charge used is in the range 20 to 80~pC, and the bunch repetition rate is
 a few hertz. We assume that when each bunch of electrons is incident on the
 screen, it produces luminescence proportional to the flux of electrons arriving at
 each point on the screen. The camera viewing the screen captures 50 images per 
 second, but is not synchronised with the arrival of the electrons at the screen.
 During the analysis of the data we find a shot-to-shot variation in the brightness
 of 10 to 20\%.

Although the ALICE tomography section was originally designed for tomographic 
measurements using three screens simultaneously, in practice it is time consuming
 to set up equal phase advances between screens. 
For this work, we have chosen to undertake the much quicker 
 quadrupole scan method. The variation of quadrupole magnet currents and the
 capture of the corresponding camera images of the screens has been automated using
 software developed in-house. In a typical measurement, the strength of QUAD-07  
 is varied and the beam images on screen 1 are captured. The quadrupole field 
 gradients are chosen to correspond to the required projection angles calculated 
 using Eq. (\ref{eq:tangent}). The equation for the transfer map between the
 entrance to QUAD-07 to screen 1 is Eq. (\ref{fig:quad7-a}) The form of this 
 function limits the angle range to about 160$^\circ$. Typically, we record images
 at 1$^\circ$ intervals, so 160 images would be collected.

\begin{figure}
\begin{center}
\subfigure[]{\label{fig:truncate-a}
\includegraphics[width=.23\textwidth]{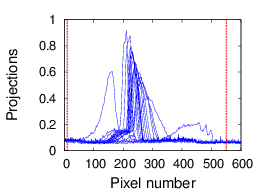}}
\subfigure[]{\label{fig:area-a}
  \includegraphics[width=.23\textwidth]{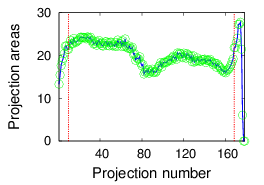}}
\end{center}
\caption{\label{fig:truncate}(a) Raw projections from the images of the QUAD-10 scan
                                    for 80 pC bunch charge.  The horizontal line
                                    through the noise floor will be used as the new zero.
							  (b) Integrated projection areas corresponding to the 
							  projection angles in Fig. \ref{fig:quad7-a}. The 
							  dashed red lines 
							  mark the region outside of which the data is also 
							  excluded because integrated areas are well below
                               the average.  
							   (The vertical axes of both graphs are in arbitrary units.)
}
\end{figure}
 
Figure \ref{fig:truncate-a} shows examples of projections obtained directly from
 the images. We call these the raw projections. Before undertaking the quadrupole
 scan measurements, a dipole magnet before the quadrupole is adjusted to centre the
 beam on the screen, so that most of the projection peaks are at roughly the same
 position. The strength of the quadrupole which we intend to use for the 
 measurements is then varied to check if the beam is also central in the magnet. 
 If the beam is on the magnetic axis of the quadrupole, it will experience no force
 and the beam spot on the screen will not move. If the beam spot moves, we adjust
 beam steering upstream of the quadrupole magnet and check again. 
 Notice for each projection that as we move away from its peak,
 the projection reaches a roughly constant, non-zero value.
The background when there is no beam has
 also been measured and found to be close to the background when the beam is 
 present. This background must be subtracted.

It is important to check the integrated area of each projection. Figure 
\ref{fig:area-a} is an example of the integrated areas calculated for a quadrupole 
scan. Note that the projection number corresponds to the projection angles in Fig. 
\ref{fig:quad7-a}, which are taken at uniform intervals. As it can be seen in the
 figure, some of the projection areas are much smaller than the typical value. This
 happens when the beam becomes defocused, but why this happens is not understood at 
 present. Including such projections could lead to errors, so they are omitted. This
 usually corresponds to the first and last few images for each of our quadrupole 
 scan data sets. In the analyses following this section, the first and last ten 
 projections are omitted, as indicated by the two vertical lines in Fig. 
 \ref{fig:area-a}. The trend in the area suggests that the bunch charge might have
 changed during the quadrupole scan. 


\section{\label{sec:space}Space Charge Search}


\subsection{\label{sec:charge}Space Charge Measurement Procedure}

There is some simulation work on the effect of different
bunch charges on the beam in ALICE \cite{Holder, DArcy, Holder2}.
These publications suggest  that at 80 pC bunch charge, 
changes in lattice functions and beamwidths become noticeable. 
If space charge effect is significant,
it would have an impact on our tomographic 
reconstruction \cite{Stratakis2}.  
In order to determine if the
space charge effect is significant,
we design an experiment as follows.  

A quadrupole scan is not by itself able to detect space charge effect.  
We propose to do it using two quadrupole scans that are separated by a 
distance that is much larger than the distance within a single scan.  
Our beam is likely to have a small space charge effect, if any.  For 
each quadrupole scan, the distance between quadrupole and screen is 
small.  Any space charge effect would be small, 
 so errors need also to be small if any effect is to be observed.
  We then do two quadrupole scans at different positions.  
As the distance between the two scans is much larger than the distance 
within each scan, the space charge effect would also be much larger.  
It is by comparing the two scans that we hope to detect the space charge 
effect.

Quadrupole scans are carried out
at screen 1 and screen 3, as shown in Fig. \ref{fig:tomo-section}.
These two screens are separated by 1.5 m.  
Using the beam images from either screen, the emittance could be obtained
as described before. 
If the space charge effect is significant, the
results from the two screens would be different.
If there is indeed no space charge effect at all, 
the phase space reconstructed from the two scans 
should also be the same.

In order to obtain reasonable reconstructions, the parameters
used in each quadrupole scan have to be selected to give
a range of projection angles as close as possible to  the full
 180$^\circ$.
The reason is that the reconstruction can in theory
be expressed as an integral of the filtered back projections
over 180$^\circ$ \cite{Kak}.  A reduced range would
in effect be a truncation in angles.
 For direct comparison, we also require that, for both scans, 
the reconstruction be carried out at the same location.

The closest quadrupole in front of screen 1 
is QUAD-07.  
We need to determine if this quadrupole could provide 
sufficient range in projection angles for the scan on screen 1.
We choose as the common reconstruction location for both scans
the entrance plane to QUAD-07.  This same quadrupole
would be used for the scan on screen 1.  Between this location
and screen 3, there are altogether five quadrupoles.
We need to decide which one to choose for the scan on screen 3.

QUAD-07 is a horizontally focussing quadrupole.  
The region between the reconstruction location
and screen 1 is made up of QUAD-07 followed by a drift space.
Using the hard-edge model for the quadrupole, we can write 
down the transfer matrix from the reconstruction location 
to screen 1:
\begin{equation}
  M =
  \left( \begin{array}{cc} 1 & L_D \\ 
                           0 & 1  \end{array} \right) 
  \left( \begin{array}{cc} \cos(\omega L) & \sin(\omega L)/\omega \\ 
                    -\omega\sin(\omega L) & \cos(\omega L) \end{array} \right) 
\label{eq:quad7}
\end{equation}
where $L$ is the quadrupole length, $L_D$ is the drift distance, and
$\omega^2$ is the normalised quadrupole field gradient 
\begin{equation}
  k_1 = \frac{e}{P_0}\frac{\partial B_y}{\partial x}.
\label{eq:k1}
\end{equation}
Here, $e$ is the electron charge,
$\partial B_y/\partial x$ is the magnetic field gradient and 
$P_0$ the electron momentum.
The magnetic field gradient has been measured as a function of
current $I$ by the manufacturer, and has
the form
\begin{equation}
  \frac{\partial B_y}{\partial x} = mI + d.
\label{eq:cal}
\end{equation}
In the case of QUAD-07, for example,
 $m = 1.5900$ T/m/A, and $d = 0.0001$ T/m.
 These values are obtained by fitting a straight line to the
 numerical data provided by the manufacturer.
 
\begin{figure}
\begin{center}
  \includegraphics[width=.35\textwidth]{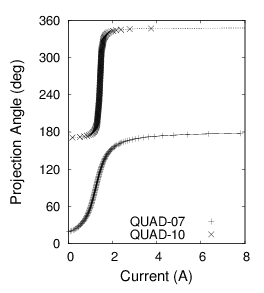}
\end{center}
\caption{\label{fig:quad7-a}  Projections angles versus QUAD-07  and QUAD-10 currents at 12 MeV.
}
\end{figure}

Using these equations, the projection angle $\theta$ can then be computed
for each current using Eq. (\ref{eq:tangent}).
A graph of the angle against current is plotted in Fig.
\ref{fig:quad7-a}.  From this graph, the range of angles can 
be obtained.

We have seen that the QUAD-07 scan for screen 1
 gives a fairly wide range of
angles, from about 20$^\circ$ to 170$^\circ$, which should
be sufficient for our purpose.  

We turn now to the scan for
screen 3.  In order to have a good, well focussed beam on the screen,
all of the quadrupoles from QUAD-07 to QUAD-11 must be on.
One of these must then be selected.  
Only QUAD-10 has a stable range that is close to 180$^\circ$.
The range at 12 MeV is plotted in Fig. \ref{fig:quad7-a}.		
There is a very steep slope that covers a large
part of the range of angles, for a small interval of currents.
This suggests that a small error in current could lead to a large
error in angle.

\subsection{\label{sec:recon}Tomographic Reconstructions}

\begin{figure}
\begin{center}
\subfigure[]{\label{fig:recon-a}
  \includegraphics[width=.23\textwidth]{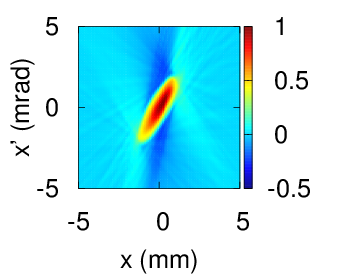}}
\subfigure[]{\label{fig:recon-b}
  \includegraphics[width=.23\textwidth]{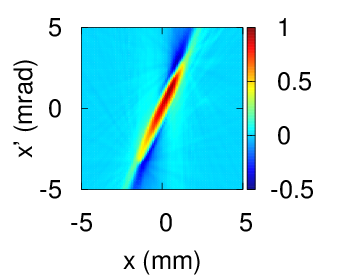}}
\subfigure[]{\label{fig:recon-c}
  \includegraphics[width=.23\textwidth]{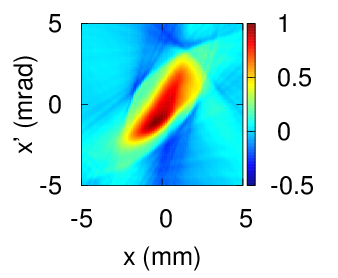}}
\subfigure[]{\label{fig:recon-d}
  \includegraphics[width=.23\textwidth]{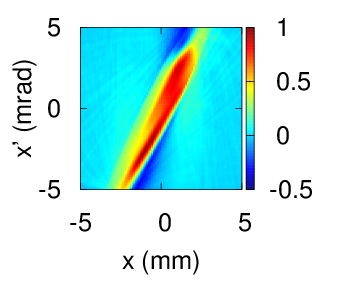}}
\end{center}
\caption{\label{fig:recon}Reconstruction at entrance face of QUAD-07 with bunch charges of: 
                          20 pC for (a) QUAD-07 and (b) QUAD-10 scans;
                          80 pC for (c) QUAD-07 and (d) QUAD-10 scans.
}
\end{figure}

The reconstructions for the QUAD-07 and QUAD-10 scans are shown 
in Fig. \ref{fig:recon} for two bunch charges, 20 pC and 80 pC.
As explained in section \ref{sec:charge},
the experiment is designed in such a way as to give 
nominally identical
reconstructions for both scans, at the entrance face to QUAD-07
- when there is no space charge effect.
Figure \ref{fig:recon-a} looks different from Fig. \ref{fig:recon-b},
and 
Fig. \ref{fig:recon-c} looks different from Fig. \ref{fig:recon-d}.
If the space charge has a linear effect, this could happen.
For instance, if the space charge defocuses the beam in the
same way as a defocussing quadrupole (both horizontally and vertically),
 this would be a linear effect.  Errors in 
quadrupole gradients and bunch to bunch variations are also
possible causes.  It is straightforward to estimate the
effects of quadrupole gradient errors, which we now do.
The estimation could be viewed as a result of either the linear
defocusing effect of space charge, or the gradient errors,
or a combination of both.

\begin{figure}
\begin{center}
\subfigure[]{\label{fig:normalised-a}
  \includegraphics[width=.23\textwidth]{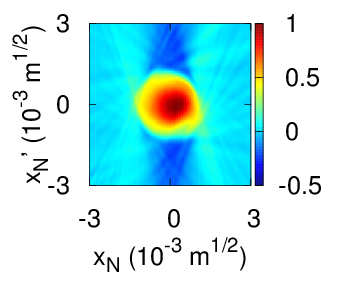}}
\subfigure[]{\label{fig:normalised-b}
  \includegraphics[width=.23\textwidth]{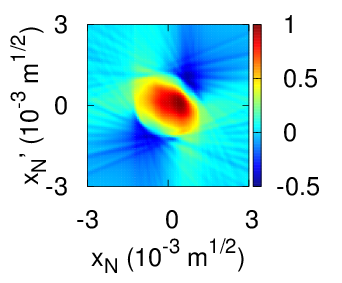}}
\subfigure[]{\label{fig:normalised-c}
  \includegraphics[width=.23\textwidth]{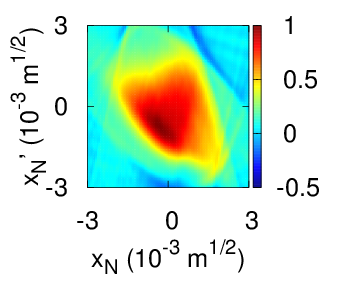}}
\subfigure[]{\label{fig:normalised-d}
  \includegraphics[width=.23\textwidth]{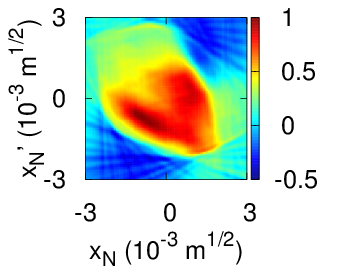}}
\end{center}
\caption{\label{fig:normalised}Normalised phase space at entrance face of QUAD-07 with bunch charges: 
                          20 pC for (a) QUAD-07 and (b) QUAD-10 scans;
                          80 pC for (c) QUAD-07 and (d) QUAD-10 scans.
}
\end{figure}

An error in the field gradient of the quadrupole could come from
an error in the current setting, or an error in the calibration
in Eq. (\ref{eq:cal}).  For our purpose, we shall combine
the two effects into a current error.  An error in the current
would lead to an error in the transfer matrix, such as
Eq. (\ref{eq:quad7}) for the QUAD-07 scan.
The result would be a reconstructed distribution that looks different
from the actual one.  However, the two distributions would be related
by a linear transfer matrix.  Assuming that current error is the cause,
if we transform both reconstructions of 
the QUAD-07 and QUAD-10 scans to normalised phase space,
the resulting distribution should look the same, 
 differing by a simple rotation at most.  
 The procedure for doing so is described in \cite{Hock}.
It requires an estimate 
 of the Twiss parameters, which are obtained
in the next section.  The resulting normalised phase space
distributions are shown in Fig. \ref{fig:normalised}.
We first summarise the procedure.

The implementation of the Filtered Back Projection 
technique normally assumes that the intervals of
angles are uniform \cite{Kak}.
In Eq. \ref{eq:BP_2}, we have given a formula 
that is suitable for nonuniform intervals of angles.  
This would be useful later, when we consider the
effect of an error in the quadrupole current.

To reconstruct in normalised phase space, we first define a
rectangular grid of the co-ordinates $(x_N, x_N')$, calculate
the corresponding co-ordinates in real space using: 
\begin{equation}
  \left( \begin{array}{c} x \\ 
                          x^\prime \end{array} \right)
= \left( \begin{array}{cc}  \sqrt{\beta}     & 0 \\ 
     -\frac{\alpha}{\sqrt{\beta}} & \frac{1}{\sqrt{\beta}} \end{array} \right)
\left( \begin{array}{c} x_N \\ 
                       x^\prime_N \end{array} \right) ,
%
\end{equation}
where $\alpha$ and $\beta$ are the Twiss parameters,
then reconstruct using Eq. (\ref{eq:BP_2}).

The structures in the phase space distributions are more
clearly visible in the normalised phase space in 
Fig. \ref{fig:normalised}, than in the real phase space in
 Fig. \ref{fig:recon}.  The structures in
 Figs. \ref{fig:normalised-a} and \ref{fig:normalised-b}
 look similar, except that \ref{fig:normalised-b} looks stretched.
 This could be due to errors in the measured Twiss parameters.
 Next, look at 
Figs. \ref{fig:normalised-c} and \ref{fig:normalised-d}.  
Both reveal similar, heart-shaped distributions,
with one rotated with respect to the other.  A
rotation is what we would expect from an error in the transfer
matrix, which could arise from errors in current or field gradient. 
As a simple test, we repeat the reconstruction of 
Fig. \ref{fig:normalised-d} from the projections.
The procedure requires the computing of the transfer matrix
from the entrance face of QUAD-07 to screen 3.
This relies on the calibration of Eq. (\ref{eq:cal}) for each
of the intervening quadrupoles.  This time, we add an error of
+0.3 A to the current settings recorded in the experiment
for QUAD-10.
The transfer matrix calculated from this new set of currents
give angles that follow the same QUAD-10 curve in
Fig. \ref{fig:quad7-a}.  A positive current error 
would cause the points to move upwards, 
suggesting that some form of rotation may take place.  

An important step has to be taken before the reconstruction can
take place correctly.
The intervals of angle for QUAD-10 in Fig. \ref{fig:quad7-a} are no longer
uniform.  This change must be applied to $\Delta \theta_k$ in
Eq. (\ref{eq:BP_2}) as weighting factors.  As demonstrated in
\cite{Hock}, a sum of half of the angular intervals on the two sides
of each projection works well.  In this set of data, we do not have
the full range of angles of 180$^\circ$.  So for the first projection,
the factor would be half of the interval between the angles of the
first two projections only.  Likewise for the last projection.
This choice maintains the equivalence of Eq. (\ref{eq:BP_2})
to the trapezium rule of integration that is explained in \cite{Hock}.

\begin{figure}
\begin{center}
\subfigure[]{\label{fig:error-a}
  \includegraphics[width=.23\textwidth]{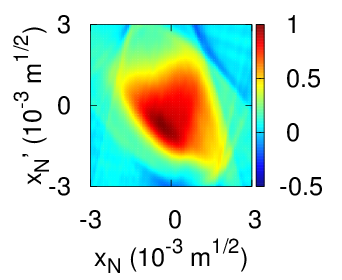}}
\subfigure[]{\label{fig:error-b}
  \includegraphics[width=.23\textwidth]{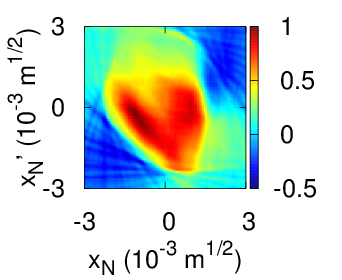}}
\end{center}
\caption{\label{fig:error}80 pC, normalised phase space: 
                         (a) reconstruction for the QUAD-07 scan;
						 (b) reconstruction for the QUAD-10 scan, assuming 
						 a current error of +0.3 A.
}
\end{figure}

Following Eqs. (\ref{eq:BP_2}) and 
(\ref{eq:actual-norm}), we reconstruct the normalised phase space
in Fig. \ref{fig:error-b}.  This is clearly rotated with respect
to the original Fig. \ref{fig:normalised-d}.  It is now at roughly
the same orientation as the screen 1 result Fig. \ref{fig:normalised-c},
reproduced in Fig. \ref{fig:error-a} for direct comparison.  
The similarity shows that the two quadrupole scans give consistent results.
As these are taken at different positions along the beamline,
the similarity also provides support that the reconstructed 
phase space distribution is correct.

This suggests that an error in quadrupole field or current could contribute to 
 the difference between the  
screen 1 result in Fig. \ref{fig:recon-c}, 
and the screen 3 result in Fig. \ref{fig:recon-d}. 
A current error 0.3 A seems rather large.
Other reasons may include fringe fields and space charge.
Further study is needed to confirm this.


\section{\label{sec:MENT_recon}MENT Reconstructions}


MENT can be used for tomographic reonstructions when the
 number of projections is small.
At ALICE, PITZ, SNS and PSI, 3 to 5 projections are used \cite{Reggiani}. 
In contrast, we could for example collect over 100 projections 
using quadrupole scans and reconstruct using FBP. 
With so few projections in MENT, it is not clear how reliable
the reconstructions are.  We review here our study \cite{Hock1} which
shows that the reconstructions are sensitive to the actual
projection angles selected and can be highly distorted, and that
by using equal angle intervals in normalised phase
space - i.e. equal phase advances - distortion can be 
reduced significantly.


\subsection{\label{sec:expt}Distortions}

\begin{figure}
\begin{center}
\subfigure[]{\label{fig:r9s-a}
  \includegraphics[width=.23\textwidth]{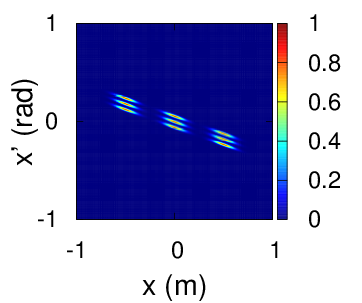}}
\subfigure[]{\label{fig:r9s-c}
  \includegraphics[width=.23\textwidth]{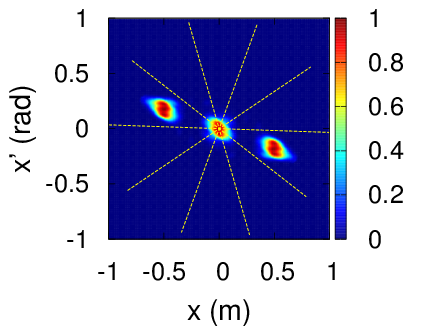}}
\subfigure[]{\label{fig:r9s-c-good}
  \includegraphics[width=.23\textwidth]{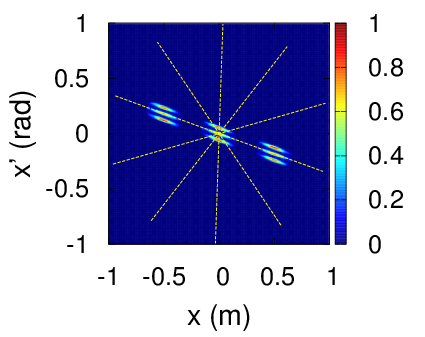}}
\subfigure[]{\label{fig:n9s-c}
  \includegraphics[width=.23\textwidth]{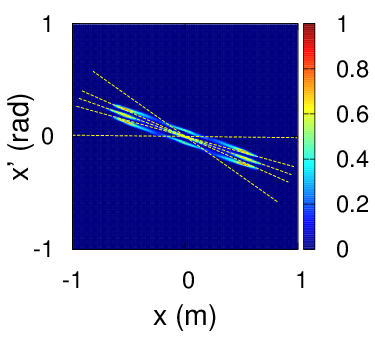}}
\subfigure[]{\label{fig:n9s-c-good}
  \includegraphics[width=.23\textwidth]{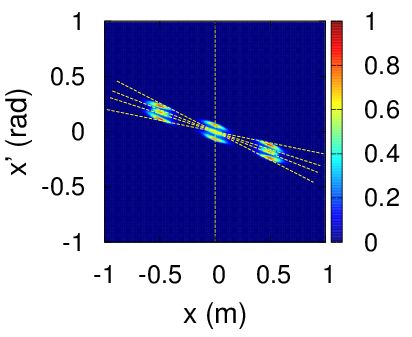}}
\end{center}
\caption{\label{fig:r9s}(a) Original distribution, with 9 spots;
						  (b) reconstructed using 5 projections
						      at equal angular intervals in real phase space, 
                                with yellow lines showing ray directions;
						  (c) the same, but with projection angles rotated half an interval; 
						  (d) reconstructed using 5 projections
						      at equal phase advances, 
                                with yellow lines showing ray directions;
						(e)	the same, but with projection angles rotated half an interval
								in normalised phase space.
}
\end{figure}

As an example of a more
complex distribution, we choose a hypothetical distribution with a number
of Gaussian spots, as shown in Fig. \ref{fig:r9s-a}.  We use this
as a test case to compare the results of the two methods for choosing
projection angles.  Figure \ref{fig:r9s-c} is the result of 
reconstructing with 5 projections at equal angular intervals in
real phase space.  The result is very 
sensitive to the actual directions of the five angles.  The result
shown here is the worst case, where the individual spots are not
resolved.  The best case shown in Fig. \ref{fig:r9s-c-good} is 
obtained when the rays are all rotated 
by half an angle interval,
actually agrees very well with the original in Fig. \ref{fig:r9s-a}.

 We now apply the method of equal phase advances, i.e.
 we use equal angles in normalised phase space.
 The result is shown in Fig. \ref{fig:n9s-c}. 
This is much closer to the original than Fig. \ref{fig:r9s-c}, 
though not as good as Fig. \ref{fig:r9s-c-good}.
Notice that when equal phase advances are chosen,
the corresponding rays in real phase space
 are closely bunched along the length of the distribution.
 This means more samples within the angular range of the 
 distribution, where it really matters.
 This is clear from the yellow lines in Fig. \ref{fig:n9s-c}.
    
If the normalised phase space angles in  
  Fig. \ref{fig:n9s-c} are changed by half an interval,
it would give   Fig. \ref{fig:n9s-c-good}.
This is slightly clearer, though still
 not as good as Fig. \ref{fig:r9s-c-good}.
So using equal phase advances give consistently reliable results,
whereas using equal angle intervals in real phase space could give 
highly distorted results for some angles.

These simulation results show that we must be careful when 
interpreting MENT results because significant distortions are 
possible.  They also provide a visual explanation for 
the conclusion that 45$^\circ$ phase advances
give minimum emittance error in the 4 screen setup
 in \cite{Castro}. It is because the angular distribution is 
 sampled optimally.
 
 
\subsection{\label{sec:expt}Re-analysing FBP Data}

Implementing equal phase advances on a beamline is possible with
some effort.  
At PITZ, equal phase advances are set up before
measurements \cite{Asova2} by adjusting upstream magnets to match
the beam distribution into the periodic Twiss parameters at the 
tomography section.
At ALICE, this set up has not been attempted.

For this analysis, we shall obtain these
phase advances in a simple way from measured data.
In our previous work at ALICE, we have reported a comprehensive set 
of phase space measurements \cite{Ibison}.  The projections are
obtained with quadrupole scans and the phase space is reconstructed
using FBP.  The basic setup consists of only one screen and one
quadrupole.  As the strength of the quadrupole is varied,
a camera captures the image on the screen repeatedly.  The procedure
is automated by a computer and each scan of the quadrupole strength
can be completed in about 10 minutes.  In a typical measurement, 
over 100 projections at 1$^\circ$ intervals are obtained. 
For this analysis, 
we simply pick out a few angles from this set of projections that
correspond to equal phase advances.  Then we reconstruct the phase space
using MENT.

Instead of having 3 to 5 screens and quite a number of quadrupoles,
 as is typical in beamlines designed to use MENT, all we need is 
 1 screen and 1 quadrupole.  It may seem redundant to use MENT for
reconstruction if we can reconstruct the phase space using
FBP.  However, there are a few good reasons:  
\begin{enumerate}
\item  A single quadrupole
cannot give the full range of projection angles \cite{Ibison},
so the FBP result tends to have streaking artefacts.
\item  MENT could produce clean results with no artefacts.  (Whether
or not it is distorted is a question we seek to answer.)
\item  Using the quadrupole scan to obtain projections
 needed for MENT is very quick and requires far less hardware
 compared to the standard procedure of using 3 to 5 screens.
\item  Having an alternative method to measure the phase space 
is useful because it provides a check for consistency.  
The MENT result could be compared with the FBP result.
\end{enumerate} 

\begin{figure}
\begin{center}
\subfigure[]{\label{fig:0347_r4-a}
  \includegraphics[width=.23\textwidth]{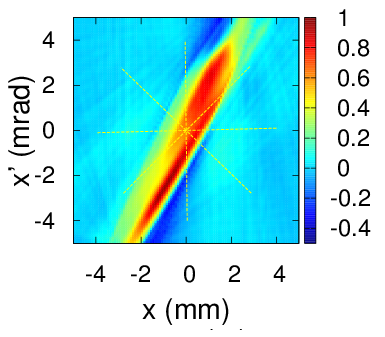}}
\subfigure[]{\label{fig:0347_r4-b}
  \includegraphics[width=.23\textwidth]{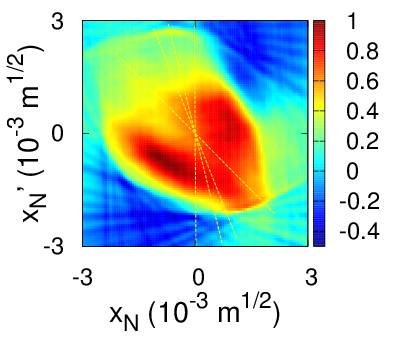}}
\subfigure[]{\label{fig:0347_r4-c}
  \includegraphics[width=.23\textwidth]{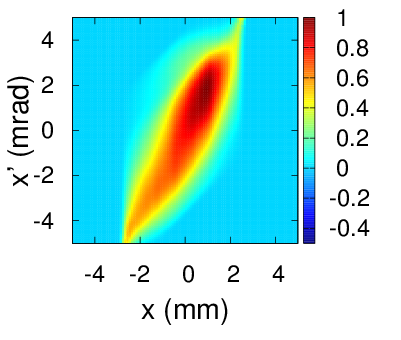}}
\end{center}
\caption{\label{fig:0347_r4}(a) FBP reconstruction, 
                                with yellow lines showing ray directions
								for equal angular intervals in real phase space;
						  (b) the same distribution and rays,
						     transformed to normalised phase space;
						  (c) reconstructed using the 4 projections
						      with MENT.
}
\end{figure}

We select experimental data from the
measurement of a beam at ALICE with 80 pC bunch charge.
The measurement setup has been reported in \cite{Ibison}.
Here, we shall assume that the projections have been measured.
The distribution has been reconstructed with FBP,
as shown in Fig. \ref{fig:0347_r4-a}.  

 With the projections from the quadrupole scan, we can  estimate 
 the Twiss parameters using the method in \cite{Ross}.  
 With a knowledge of the Twiss parameters,
 we can then transform the distribution to normalised phase space,
 as shown in Fig. \ref{fig:0347_r4-b}.  
To apply MENT, we first try it for the case of projections with equal 
 angular intervals.
 We pick out 4 angles, as shown by the yellow lines in Fig. \ref{fig:0347_r4-a}.
 We must be careful to skip over the gap that is not
 covered by the  range of projection angles that is possible with a single quadrupole.  
 The corresponding rays in normalised phase space are shown 
 by the yellow lines in Fig. \ref{fig:0347_r4-b}.  They are now 
 bunched into a small range of angles.  Applying MENT to these projections,
 we get Fig. \ref{fig:0347_r4-c}.  This is clearly broader
and apparently distorted when compared with  Fig. \ref{fig:0347_r4-a}.
However, we should reserve judgement at this stage because we know that
 Fig. \ref{fig:0347_r4-a}  is also not perfect.
 
\begin{figure}
\begin{center}
\subfigure[]{\label{fig:0347_n4-a}
  \includegraphics[width=.23\textwidth]{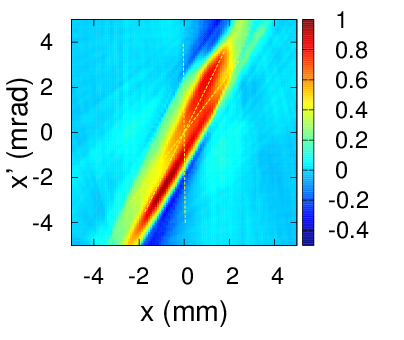}}
\subfigure[]{\label{fig:0347_n4-b}
  \includegraphics[width=.23\textwidth]{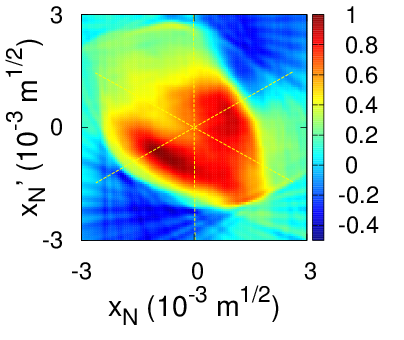}}
\subfigure[]{\label{fig:0347_n4-c}
  \includegraphics[width=.23\textwidth]{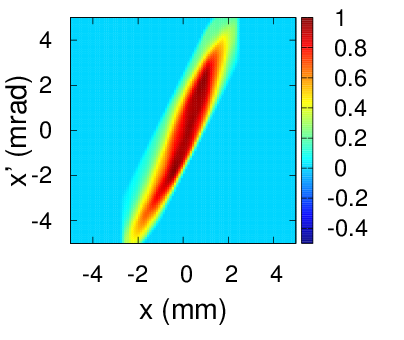}}
\end{center}
\caption{\label{fig:0347_n4}(a) FBP reconstruction, 
                                with yellow lines showing ray directions
								for equal phase advances;
						  (b) the same distribution and rays,
						     transformed to normalised phase space;
						  (c) reconstructed using the 4 projections
						      with MENT.
}
\end{figure}

 Next, we apply the method of equal phase advances.  We know from
 \cite{Hock} that this means equal angles in normalised phase space.
 So we pick four angles in normalised phase space, as shown by the 
yellow lines in Fig. \ref{fig:0347_n4-b}.
Again, we must be careful to skip over the gap in the angular range.
(If the gap is too large, fewer projections would be possible
and the experiment might have to be redesigned.
This could mean changing the quadrupole's strength and its distance
from the screen to increase the range of projection angles.)
The corresponding angles in real phase space are shown by yellow lines
in  Fig. \ref{fig:0347_n4-a}.  Notice that they are bunched closer to
the length of the FBP distribution.  The projections are reconstructed
with MENT.  The result in  Fig. \ref{fig:0347_n4-c} clearly shows
better agreement with the FBP result than Fig. \ref{fig:0347_r4-c}.

This demonstration provides support for the 
the method of equal phase advance.  It also suggests that quadrupole
scan is a possible setup in which we could use MENT with
 the method of equal phase advance.





\section{\label{sec:conclude}Conclusion}


We have presented a coherent view of the normalised phase space
method for phase space tomography:

\begin{enumerate}

\item   In 2003, a method to measure emittance at the Tesla Test Facility 2
        is developed \cite{Castro}.  The method uses 4 screens.  Adjacent screens are
		separated by identical FODO cells.  Simulations show that statistical
		errors in emittance measurements are minismised by choosing
		a setup in which phase advance is 45$^\circ$ between adjacent screens.
		
\item   In 2007, this idea is used to design the PITZ tomography section \cite{Asova}.
        The 45$^\circ$ value is now associated with 180$^\circ$ divided by 4,
		the number of screens.  The 180$^\circ$ is in turn associated with 
		the full range of projection angles in a tomographic measurement.  
		The reason for this association is not explained, but the use of 45$^\circ$
		phase advance is justified using the emittance measurement method as in 
		\cite{Castro}.  In this way, the idea that equal phase advances is optimal
		for tomographic measurement is first proposed.
				
\item   In 2008, the idea of equal phase advance is applied to the design of the ALICE
        tomography section \cite{Muratori}.  This time, the idea is used directly without
        the justification of emittance measurement.  		

\item   In 2010, both the PITZ tomography section \cite{Asova2} and the ALICE 
        tomography 
        section \cite{Muratori2} are commissioned. Subsequent tomographic 
		measurements at 
		PITZ has followed closely a procedure to setup equal phase advances 
		\cite{Asova2}.
		At the ALICE, beam time dedicated to tomographic measurement has been 
		limited.
        Instead, quick quadrupole scans are used without any phase advance setup.		        		

\item   In 2011, we supply the justification for equal phase advance
        by showing that it is equal to the projection angle in a normalised phase
		space \cite{Hock}.  In this phase space, the distribution is 
		circular on average, and equal intervals of projection angles become an optimal
        choice.  		

\item   Since then, we have applied the idea of equal phase advances to improve 
        resolution in FBP
        reconstructions \cite{Hock}, detect linear errors in a beamline 
		\cite{Ibison} and
        improve reliability in MENT reconstructions \cite{Hock1}.  We plan to 
		apply this to improve resolution 
        and reliability in 4D reconstruction where the number of
		projections that can be measured
         is likely to be limited by measurement time \cite{Hock2}.		

\end{enumerate}




\begin{acknowledgments}
We would like to thank Ben Shepherd, Rob Smith, Nino Cutic and Georgios Kourkafas
 for helpful discussion.
We are grateful to the Science and Technology Facilities Council, UK,
for financial support.
\end{acknowledgments}






\end{document}